\documentstyle[multicol,aps,epsf,amssymb]{revtex}

\newcommand{\past}{\stackrel{\leftarrow}{S}}

\newcommand{\future}{\stackrel{\rightarrow}{S}}
\newcommand{\all}{\stackrel{\leftrightarrow}{S}}

\def\Pr   {{\rm Pr}}

\def\hmu   {h_\mu}

\def\Abet {{\cal A}}
\def\EE   {{\bf E} }
\def\SI   {{\bf S} }
\def\TP   {{\bf G}}
\def\TI   {{\bf T} }
\def\l2   {{\rm log}_2}

\newtheorem{theorem}{Theorem}
\newtheorem{definition}{Definition}
\newtheorem{proposition}{Proposition}
\newtheorem{lemma}{Lemma}
\newtheorem{corollary}{Corollary}

\begin{document}


\title{Regularities Unseen, Randomness Observed:\\
Levels of Entropy Convergence}

\author{James P. Crutchfield}
\address{Santa Fe Institute, 1399 Hyde Park Road, Santa Fe, NM 87501\\
Electronic Address: chaos@santafe.edu}

\author{David P. Feldman}
\address{College of the Atlantic, 105 Eden St., Bar Harbor, ME 
04609\\and Santa Fe Institute, 1399 Hyde Park Road, Santa Fe, NM 87501\\
Electronic Address: dpf@santafe.edu}

\date{\today}
\maketitle

\bibliographystyle{unsrt}

\begin{abstract}
We study how the Shannon entropy of sequences produced by an
information source converges to the source's entropy rate. We
synthesize several phenomenological approaches to applying information
theoretic measures of randomness and memory to stochastic and
deterministic processes by using successive derivatives of the Shannon
entropy growth curve. This leads, in turn, to natural measures of
apparent memory stored in a source and the amounts of information that
must be extracted from observations of a source in order for it to be
optimally predicted and for an observer to synchronize to it. One
consequence of ignoring these structural properties is that the missed
regularities are converted to apparent randomness. We demonstrate that
this problem arises particularly for small data sets; e.g., in
settings where one has access only to short measurement sequences.

PACS: 02.50.Ey  
05.45.-a  
05.45.Tp  
89.75.Kd  
~~~~~~~~~~Santa Fe Institute Working Paper 01-02-012

\end{abstract}



\begin{multicols}{2}

\tableofcontents


\section{Introduction}

\subsection{Apparent Randomness}

Natural processes appear unpredictable to varying degrees and for
several reasons. First, and most obviously, one may not know the
``rules'' or equations that govern a particular system.  That is, an
observer may have only incomplete knowledge of the forces controlling
a process.  Laplace was well aware of these sources of apparent
randomness; as he commented two centuries ago in motivating his
{\em Philosophical Essay on Probabilities} 
\cite{Lapl76a}:
\begin{quotation}
But ignorance of the different causes involved in the production of
events, ...  taken together with the imperfection of analysis,
prevents our reaching the same certainty about the vast majority of
phenomena. Thus there are things that are uncertain for us, things
more or less probable, and we seek to compensate for the impossibility
of knowing them by determining their different degrees of likelihood.
\end{quotation}

Second, there may be mechanisms intrinsic to a process that amplify
unknown or uncontrolled fluctuations to unpredictable macroscopic
behavior.  Manifestations of this sort of randomness include {\em
deterministic chaos} and {\em fractal separatrix} structures bounding
different basins of attraction.  As Poincar\`e noted \cite{Poin52a}:
\begin{quotation} 
... it may happen that small differences in the initial conditions  
produce very great ones in the final phenomena. A small error in
the former will produce an enormous error in the latter. Prediction
becomes impossible, and we have the fortuitous phenomenon.
\end{quotation}
Unpredictability of this kind also arises from sensitive dependence
on parameters, such as that seen in non-structurally stable systems
with continuous bifurcations \cite{Newh81a} or from sensitive
dependence on boundary conditions. Knowledge of the governing equations
of motion does little to make these kinds of intrinsic randomness go
away.  

Third, and more subtly, there exists a wide array of observer-induced
sources of apparent randomness.  For one, the choice of representation
used by the observer may render a system unpredictable. For example,
representing a square wave in terms of sinusoids requires specifying
an infinite number of amplitude coefficients.  Truncating the order of
approximation leads to errors, even for a source as simple and
predictable as a square wave.  Similarly, an observer's choice and
design of its measuring instruments is an additional source of
apparent randomness.  As one example, Ref.~\cite{Crut91e} shows how
irreducible unpredictability arises from a measurement instrument's
distortion of a spatio-temporal process's internal states.  

Fourth, the measurement process engenders apparent randomness in
other, perhaps more obvious ways, too. Even if one knows the equations 
of motion governing a system, accurate prediction may not be possible:
the measurements made by an observer may be inaccurate, or, if the
measurements are precise, there may be an insufficient volume of
measurement data.  Or, one may simply not have a sufficiently long
measurement stream, for example, to disambiguate several internal
states and, therefore, their individual consequences for the process's
future behavior cannot be accurately accounted for.  Examples of these
sorts of measurement-induced randomness are considered in
Refs.~\cite{Bial00a,Neme00a,Schu96}.  In all of these cases, the
result is that the process appears more random than it actually is. 

Fifth, and finally, if the dynamics are sufficiently complicated it may
simply be too computationally difficult to perform the calculations
required to go from measurements of the system to a prediction of the
system's future behavior.  The existence of deeply complicated dynamics
for which this was a problem was first appreciated by Poincar\'e more
than a century ago as part of his detailed analysis of the three-body
problem \cite{Poin92}.

Of course, most natural phenomena involve, to one degree or another,
almost all of these separate sources of ``noise''.  Moreover, the
different mechanisms interact with each other. It is no surprise,
therefore, that describing and quantifying the degree of a process's
apparent randomness is a difficult yet essential endeavor that cuts
across many disciplines.   

\subsection{Untangling the Mechanisms}

A central goal here is to examine ways to untangle the different
mechanisms responsible for apparent randomness by investigating
several of their signatures. As one step in addressing these issues,
we analyze those aspects of apparent randomness over which an observer
may have some control.  These include the choice of how to quantify
the degree of randomness (e.g., through choices of statistic or in
modeling representation) and how much data to collect.  We describe
the stance taken by the observer toward the process to be analyzed in
terms of the {\em measurement channel } --- an adaptation
\cite{Crut83c} of Shannon's notion of a communication channel.  One of
the central questions addressed in the following is, how does an
observer, apprised of a process's possible states and its dynamics,
come to know in what internal state the process is? We will show that 
this is related to another question, How does an observer come to
accurately estimate how random a source is?  In particular, we shall
investigate how finite-data approximations converge to this asymptotic
value.  We shall see a variety of different convergence behaviors and
will present several different quantities that capture the nature of
this convergence.  As the title of this work suggests, we shall see
that regularities that are unseen are ``converted'' to apparent
randomness.  

It is important to emphasize, and this will be clear through our
citations, that much of our narrative about levels of entropy
convergence touches on and restates results and intuitions known to a
number of researchers in information theory, dynamical systems,
stochastic processes, and symbolic dynamics. Our attempt here, in
light of this, is several-fold.  First, we put this knowledge into a
single framework, using the language of discrete derivatives and
integrals.  We believe this approach unifies and clarifies a number of
extant quantities.  Second, and more importantly, by considering
numerous examples, we shall see that examining levels of entropy
convergence can give important clues about the computational structure 
of a process.  Finally, our view of entropy convergence will lead
naturally to a new quantity, the {\em transient information} ${\bf
T}$. We shall prove that the transient information captures the total
uncertainty an observer must overcome in synchronizing to a Markov
process. 

We begin in Sec.~\ref{Information_Theory} by fixing notation and
briefly reviewing the motivation and basic quantities of information
theory.  In Sections \ref{Levels_of_Convergence} and
\ref{Convergence_Integrals} we use discrete derivatives and integrals
to examine entropy convergence.  In so doing, we recover a number of
familiar measures of randomness, predictability, and ``complexity''.
Then, in Sec.~\ref{T_Section} we introduce, motivate, and interpret a
new information theoretic measure of structure, the transient
information. In particular, we shall see that the transient
information provides a quantitative measure of the manner in which an
observer synchronizes to a source.  We then illustrate the utility of
the quantities discussed in
Sec.~\ref{Levels_of_Convergence}--\ref{T_Section} by considering a
series of increasingly rich examples in
Sec.~\ref{Examples_Section}. In Sec.~\ref{Applications} we look at
relationships between the quantities discussed previously.  In 
particular, we show several quantitative examples of how regularities
that go undetected are converted into apparent randomness.  Finally, we
conclude in Sec.~\ref{Conclusions} and offer thoughts on possible
future directions for this line of research. 


\section{Information Theory}
\label{Information_Theory}
\subsection{The Measurement Channel}

In the late 1940's Claude Shannon founded the field of communication
theory \cite{Shan48}, motivated in part by his work in cryptography
during World War II \cite{Shan93a}. His attempt to analyze the basic
trade-offs in disguising information from third parties in ways that
still allowed recovery by the intended receiver led to a study of how
signals could be compressed and transmitted efficiently and error
free. His basic conception was that of a {\em communication channel}\/
consisting of an {\em information source}\/ which produces {\em
messages} \/ that are encoded and passed through a possibly noisy and
error-prone channel.  A {\em receiver} then decodes the channel's
output in order to recover the original messages. Shannon's main
assumptions were that an information source was described by a
distribution over its possible messages and that, in particular, a
message was ``informative'' according to how surprising or unlikely 
its occurrence was. 

We adapt Shannon's conception of a communication channel as follows:
We assume that there is a {\em process} \/ (source) that produces a
{\em data stream} \/(message) --- an infinite string of symbols drawn
from some finite alphabet.  The task for the {\em observer} \/
(receiver) is to estimate the probability distribution of sequences
and, thereby, estimate how random the process is.  Further, we
assume that the observer does not know the process's structure; the
range of its states and their transition structure --- the process's
internal dynamics --- are hidden from the observer.  (We will,
however, occasionally relax this assumption below.)  Since the
observer does not have direct access to the source's internal, hidden
states, we picture instead that the observer can estimate to arbitrary
accuracy the probability of measurement sequences.  Thus, we do not
address the eminently practical issue of how much data is required for
accurate estimation of these probabilities. For this see, for example, 
Refs.~\cite{Schu96,Shaw84,Gras86}. In our scenario, the observer
detects sequence blocks directly and stores their probabilities as
histograms. Though apparently quite natural in this setting, one
should consider the histogram to be a particular class of
representation for the source's internal structure --- one that may or
may not correctly capture that structure.  

\begin{figure}[tbp]
\epsfxsize=3.0in
\begin{center}
\leavevmode
\epsffile{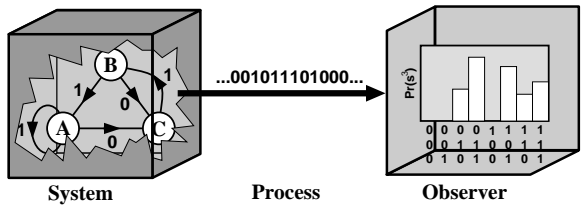}
\end{center}
\caption{The measurement channel: The internal states
$\{{\bf A}, {\bf B}, {\bf C} \}$ of the system are reflected, only
indirectly, in the observed measurement of $1$'s and $0$'s. An observer
works with this impoverished data to build a model of the underlying
system. After Ref.~\protect\cite{Crut91b}.}
\label{MeasurementChannel}
\end{figure}

This {\em measurement channel} \/ scenario is illustrated in
Fig.~\ref{MeasurementChannel}.  In this case, the source is a
three-state deterministic finite automaton. However, the observer does 
not see the internal states $\{ {\bf A}, {\bf B}, {\bf C}
\}$. Instead, it has access to only the measurement symbols $\{ 0, 1
\}$ generated on state-to-state transitions by the hidden automaton.
In this sense, the measurement channel acts like a communication
channel; the channel maps from a internal-state sequence $\ldots {\bf
BCBAACBC} \ldots$ to a measurement sequence $\ldots 0111010 \ldots$.
The process shown in Fig.~\ref{MeasurementChannel} belongs to the
class of stochastic process known as {\em hidden Markov models}.  The
transitions from internal state to internal state are Markovian, in
that the probability of a given transition depends only upon which
state the process is currently in. However, these internal states are
not seen by the observer --- hence the name ``hidden'' Markov model
\cite{Blac57a,Elli95}. 

Given this situation, a number of issues arise for the observer.  One
fundamental question is how many of the system's properties can be
inferred from the observed binary data stream. In particular, can the
observer build a model of the system that allows for accurate
prediction?  According to Shannon's coding 
theorem, success in answering these questions depends on whether the
system's entropy rate falls below the measurement channel capacity. If
it does, then the observer can build a model of the
system. Conversely, if the entropy rate is above the
measurement channel's capacity, then the theorem tells us that the
observer cannot exactly reconstruct all properties of the system. In
this case, source messages --- sequences over internal states ---
cannot be decoded in an error-free manner. In particular, optimal
prediction will not be possible. In the following, we assume that the
channel capacity is larger than the entropy rate and, hence, that optimal
prediction is --- in theory, at least --- possible.  

Similar questions of building models from data produced by various
kinds of information sources are found in the fields of machine
learning and computational learning theory. See the appendices in
Ref.~\cite{Shal98a} for comments on the similarities and differences
with the approach taken here.

\subsection{Stationary Stochastic Processes}

The measurement streams we shall consider will be stationary
stochastic processes.  In this section we introduce this idea more
formally, fix notation, and define a few classes of stochastic
process to which we shall return when considering examples in
Sec.~\ref{Examples_Section}. 

The main object of our attention will be a one-dimensional chain
\begin{equation}
\stackrel{\leftrightarrow}{S} \, \equiv \,\ldots S_{-2} S_{-1} S_0 S_1
  \ldots 
\end{equation}
of random variables $S_t$ that range over a finite set $\Abet$. We
assume that the underlying system is described by a shift-invariant
measure $\mu$ on infinite sequences
$\cdots s_{-2} s_{-1} s_0 s_1 s_2 \cdots; s_t \in \Abet$
\cite{Gray90a}. The measure $\mu$ induces a family of distributions, 
$\{ {\rm Pr}(s_{t+1} , \ldots , s_{t+L}): s_t \in \Abet \}$, where
${\rm Pr}(s_t)$ denotes
the probability that at time $t$ the random variable $S_t$ takes
on the particular value $s_t \in \Abet$ and ${\rm Pr} (s_{t+1} ,
\ldots , s_{t+L})$ denotes the joint probability over blocks of $L$
consecutive symbols.  We assume that the distribution is stationary;
${\rm Pr}(s_{t+1},\ldots, s_{t+L})={\rm Pr}(s_1, \ldots , s_L )$. 

We denote a block of $L$ consecutive variables by $S^L \equiv S_1
\ldots S_L$.  We shall follow the convention that a capital letter
refers to a random variable, while a lowercase letter denotes a
particular value of that variable.  Thus, $s^L = s_0 s_1 \cdots
s_{L-1}$, denotes a particular symbol block of length $L$.
We shall use the term {\em process} to refer to the joint distribution
${\rm Pr} (\stackrel{\leftrightarrow}{S}) $ over the infinite chain of
variables.  A process, defined in this way, is what Shannon referred
to as an information source.

For use later on, we define several types of processes.  First, and
most simply, a process with a {\em uniform distribution}\/ is one in
which all sequences occur with equiprobability.  We will denote this
distribution by $U^L$;
\begin{equation}
U(s^L) \, = \, 1 / | {\cal A} |^{L} \;.
\end{equation}
Next, a process is {\em independently and identically distributed} (IID)
if the joint distribution ${\rm Pr}(\stackrel{\leftrightarrow}{S}) =
{\rm Pr}( \ldots, S_i, S_{i+1}, S_{i+2}, S_{i+3}, \ldots )$ factors in
the following way: 
\begin{equation}
{\rm Pr}(\stackrel{\leftrightarrow}{S}) \, = \, \ldots {\rm
 Pr}(S_i) {\rm Pr}(S_{i+1}) {\rm Pr}(S_{i+2}) \ldots \;,
\label{IID}
\end{equation}
and ${\rm Pr}(S_i)$ = ${\rm Pr}(S_j)$ for all $i, j$. 

We shall call a process {\em Markovian}\/ if the probability of the
next symbol depends only on the previous symbol seen.  In other words,
the joint distribution factors in the following way:
\begin{equation}
{\rm Pr}(\stackrel{\leftrightarrow}{S}) =
  \ldots {\rm Pr}(S_{i+1} | S_i) {\rm Pr}(S_{i+2} | S_{i+1}) \ldots 
\end{equation}
More generally, a process is {\em order-$R$ Markovian}\/ if the
probability of the next symbol depends only on the previous $R$
symbols:
\begin{equation}
{\rm Pr}( S_i | \ldots, S_{i-3}, S_{i-1}) =
 {\rm Pr}( S_i | S_{i-R}, \ldots, S_{i-1})\;. 
\label{OrderR.Markovian.def}
\end{equation} 
Finally, a {\em hidden Markov process} consists of an internal order-$R$
Markov process that is observed only by a function of its internal-state
sequences. These are sometimes called {\em functions of a Markov chain}
\cite{Blac57a,Elli95}. We refer to all of these processes as {\em
finitary}, since there is a well defined sense, discussed below, in
which they have a finite amount of memory.

\subsection{Basic Quantities of Information Theory} 

Here, we briefly state the definitions and interpretations of the basic
quantities of information theory.  For more
details, see Ref.~\cite{Cove91}. Let $X$ be a random variable that
assumes the values $x \in {\cal X}$, where ${\cal X}$ is a finite set.
We denote the probability that $X$ assumes the particular value $x$ by
${\rm Pr}(x)$.  Likewise, let $Y$ be a random variable that assumes the
values $y \in {\cal Y}$.  

The {\em Shannon entropy} of $X$ is defined by:
\begin{equation}
  H[X] \equiv - \sum_{x \in {\cal X}} {\rm Pr}(x) \log_2 {\rm
  Pr}(x) \;. 
\end{equation}
Note that $H[X] \geq 0$.  The units of $H[X]$ are {\em bits}.
The entropy $H[X]$ measures the uncertainty associated with the random
variable $X$.  Equivalently, it measures the average amount of memory,
in bits, needed to store outcomes of the variable $X$.  
The {\em conditional entropy} is defined by
\begin{equation}
 H[X|Y] \equiv - \sum_{x \in {\cal X}, y \in {\cal Y}} {\rm
  Pr}(x,y)  \log_2 {\rm Pr}(x|y) \;,
\end{equation}
and measures the average uncertainty associated with variable $X$,
if we know $Y$.

The {\em mutual information} between $X$ and $Y$ is defined as
\begin{equation}
  I[X;Y] \equiv H[X] - H[X|Y] \;.
\label{mutual.info.ent.diff}
\end{equation}
In words, the mutual information is the average reduction of
uncertainty of one variable due to knowledge of another.  If knowing
$Y$ on average makes one more certain about $X$, then it makes sense
to say that $Y$ carries information about $X$.  Note that $I[X;Y] \geq
0$ and that $I[X;Y] = 0$ when either $X$ and $Y$ are independent
(there is no ``communication'' between $X$ and $Y$) or when either
$H[X] = 0$ or $H[Y] = 0$ (there is no information to share).  Note
also that $I[X;Y] = I[Y;X]$.

The {\em information gain} between two distributions ${\rm Pr}(x)$ and
$\widehat{{\rm Pr}}(x)$ is defined by:
\begin{equation}
  {\cal D}[{\rm Pr}(x)|| \widehat{{\rm Pr}}(x)] \equiv
  \sum_{x \in {\cal X}} {\rm Pr}(x)
  \log_2 \frac{ {\rm Pr}(x)}{ \widehat{{\rm Pr}}(x)} \;, 
\label{info.gain.def}
\end{equation}
where $\widehat{{\rm Pr}}(x) = 0$ only if ${\rm Pr}(x) =
0$. Quantitatively, ${\cal D}[P||Q]$ is the number of bits by which
the two distributions $P$ and $Q$ differ \cite{Cove91}.  Informally,
${\cal D}[P||Q]$ can be viewed as the distance between $P$ and $Q$ in
a space of distributions.  However, ${\cal D}[P||Q]$ is not a metric,
since it does not obey the triangle inequality.

Similarly, the {\em conditional entropy gain} between two conditional
distributions  ${\rm Pr}(x|y)$ and $\widehat{{\rm Pr}}(x|y)$ is defined by:
\begin{equation}
   {\cal D}[{\rm Pr}(x|y)||\widehat{{\rm Pr}}(x|y)] \equiv
   \sum_{x \in {\cal X}, y \in {\cal Y} } {\rm Pr}(x,y)
   \log_2 \frac{ {\rm Pr}(x|y)}{ \widehat{{\rm Pr}}(x|y)} \;, 
\label{cond.info.gain.def}
\end{equation}

\subsection{Block Entropy and Entropy Rate}

We now examine the behavior of the Shannon entropy $H(L)$ of ${\rm
Pr}(s^L)$, the distribution over blocks of $L$ consecutive
variables.  We shall see that examining how the Shannon entropy of a
block of variables grows with $L$ leads to several quantities that
capture aspects of a process's randomness and different features of
its memory.  

The {\em total Shannon entropy} of length-$L$ sequences is defined
\begin{equation}
H(L) \, \equiv \, - \sum_{ s^L \in {\cal A}^L } \Pr (s^L) \l2 \Pr
(s^L) \;, 
\end{equation}
where $L > 0$. The sum is understood to run over all possible
blocks of $L$ consecutive symbols.  If no measurements are made, there
is nothing about which to be uncertain and, thus, we define $H(0)
\equiv 0$.  Below we will show that $H(L)$ is a non-decreasing
function of $L$; $H(L) \geq H(L\!-\!1)$.  We shall also see that it is
concave; $H(L) - 2H(L\!-\!1) + H(L\!-\!2) \leq 0$.  

\begin{figure}[tbp]
\epsfxsize=2.4in
\begin{center}
\leavevmode
\epsffile{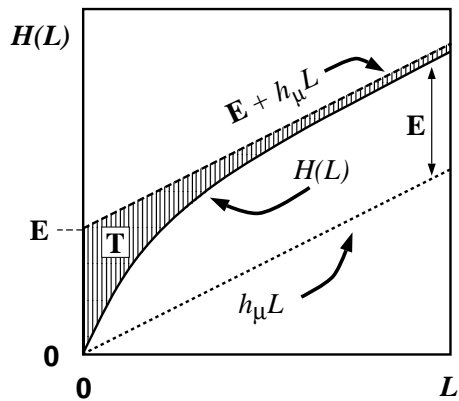}
\end{center}
\caption{Total Shannon entropy growth for a finitary information
source: a schematic plot of $H(L)$ versus $L$. $H(L)$ increases
monotonically and asymptotes to the line $\EE + h_\mu L$, where $\EE$
is the excess entropy and $h_\mu$ is the source entropy rate. This
dashed line is the $\EE$-memoryful Markovian source approximation to a
source with entropy growth $H(L)$. The entropy growth of the
memoryless-source approximation of the source is indicated by the
short-dashed line $\hmu L$ through the origin with slope $\hmu$. The
shaded area is the transient information $\TI$.  For more discussion,
see text.} 
\label{HvsL}
\end{figure}

Note that the maximum average information per observation is $\log_2 |
\Abet |$, $H(1) \leq \l2 | \Abet |$, and, more generally,
\begin{equation}
H(L) \leq L ~ \l2 | \Abet | \;.
\label{block.inequality}
\end{equation}
Equality in Eq.~(\ref{block.inequality}) occurs only when the
distribution over $L$-blocks is uniform; i.e., given by $U^L$.  Figure
\ref{HvsL} shows $H(L)$ for a typical information source.  The various
labels and the interpretation of $H(L)$ there will be discussed fully
below. 

The {\em source entropy rate} $\hmu$ is the rate of increase with
respect to $L$ of the total Shannon entropy in the large $L$ limit:
\begin{equation}
    \hmu \equiv \lim_{L \rightarrow \infty} \frac{H(L)}{L} \; ,
\label{ent.def}
\end{equation}
where $\mu$ denotes the measure over infinite sequences that induces 
the $L$-block joint distribution ${\rm Pr} (s^L)$; the units are
{\em bits/symbol}.  The limit in Eq.~(\ref{ent.def}) exists for all
stationary measures $\mu$ \cite{Cove91}.  The entropy rate $h_\mu$
quantifies the irreducible randomness in sequences produced by a
source: the randomness that remains after the correlations and
structures in longer and longer sequence blocks are taken into
account. The entropy rate is also known as the {\em thermodynamic
entropy density} in statistical mechanics or the {\em metric entropy}
in dynamical systems theory.  

As Shannon proved in his original work, $h_\mu$ also measures the
length, in bits per symbol, of the optimal, uniquely decodable, binary
encoding for the measurement sequence. That is, a message of $L$
symbols requires (as $L \rightarrow \infty$) only $\hmu L$ bits of
information rather than $\l2 | \Abet | L$ bits. This is consonant with
the idea of $h_\mu$ as a measure of randomness. On the one hand, a
process that is
highly random, and hence has large $h_\mu$, is difficult to compress.
On the other hand, a process with low $h_\mu$ has many correlations
between symbols that can be exploited by an efficient coding scheme.   

As noted above, the limit in Eq.~(\ref{ent.def}) is guaranteed to
exist for all stationary sources.  In other words,
\begin{equation}
H(L) \sim \hmu L \;\; {\rm as} \; L \rightarrow \infty \;.  
\label{SimpleScalingForm}
\end{equation}
However, knowing the value of $h_\mu$ indicates nothing about how
$H(L)/L$ approaches this limit. Moreover, there may be --- and indeed
usually are --- sublinear terms in $H(L)$. For example, one may have
$H(L) \sim c + \hmu L$ or $H(L) \sim \log L + \hmu L$.  We shall see
below that the sublinear terms in $H(L)$ and the manner in which
$H(L)$ converges to its asymptotic form reveal important structural
information about a process.

\subsection{Redundancy}
\label{RedundancySection}

Before moving on to our main task --- considering what can be learned
from looking at the entropy growth curve $H(L)$ --- we introduce one
additional quantity from information theory.
Since we are using an alphabet of size $| \Abet |$, if nothing else is
known about the process or the channel we can consider the
measurement channel used to observe the process to have a {\em channel
capacity} of ${\cal C} = \l2 | \Abet |$.  Said another way, the
maximum observable entropy rate for the channel output (the measurement
sequence) is $\log_2 | \Abet |$.  

Frequently, however, the observed $h_\mu$ is less than its maximum
value.  This difference is measured by the {\em redundancy} \/ $\bf
R$: 
\begin{equation}
{\bf R} \equiv \l2 | \Abet | - \hmu \;.
\label{Redundancy}
\end{equation}
Note ${\bf R} \geq 0$.
If ${\bf R} > 0$, then the series of random variables
$\ldots, S_i, S_{i+1}, \ldots$ has some degree of regularity: either
the individual variables are biased in some way or there are
correlations between them. Recall that the entropy rate measures the
size, in bits per symbol, of the optimal binary compression of the
source.  The redundancy, then, measures the amount by which a given
source can be compressed.  If a system is highly redundant, it can be
compressed a great deal.  

For another interpretation of the redundancy, one can show that ${\bf
R}$ is the information gain of the source's actual distribution ${\rm
Pr}(s^L)$ with respect to the uniform distribution ${\rm U}(s^L)$ in
the $L \rightarrow \infty$ limit:   
\begin{equation}
{\bf R} = \lim_{L \rightarrow \infty}
   \frac {{\cal D} [ {\rm Pr}(s^L) || {\rm U}(s^L) ] } {L}\;,
\end{equation}
where $\cal D$ is defined in Eq.~(\ref{info.gain.def}).
Restated, then, the redundancy ${\bf R}$ is a measure of the
information gained when an observer, expecting a uniform
distribution, learns the actual distribution over the sequence.

\section{Levels of Entropy Convergence: Derivatives of $H(L)$}
\label{Levels_of_Convergence}
\label{ConvergenceHierarchy}

With these preliminaries out of the way, we are now ready to begin
the main task: examining the growth of the entropy curve $H(L)$.
In particular, we shall look carefully at the manner in which the
block entropy $H(L)$ converges to its asymptotic form --- an issue that
has occupied the attention of many researchers
\cite{Bial00a,Neme00a,Shaw84,Gras86,Gatl72,Crut83a,Gyor85,Szep86,Lind88b,Szep89a,Csor89a,Li91,Crut82b,Crut82c,Pack82,Freu96a,Freu96b,Lind88a,Lind89a,Kauf91a,Csor93b,Ebel94c,Ebel97b,Crut97a,Feld98a,Ebel99a}.
In what follows, we present a systematic method for examining entropy
convergence.  To do so, we will take discrete derivatives of $H(L)$
and also form various integrals of these derivatives.  This method
allows one to recover a number of quantities that have been introduced
some years ago and that can be interpreted as different aspects of a
system's memory or structure.  Additionally, our discrete derivative
framework will lead us to define a new quantity, the {\em transient
information}, which may be interpreted as a measure of how difficult
it is to synchronize to a source, in a sense to be made precise below. 

Before continuing, we pause to note that the representation shown in the
entropy growth curve of Fig.~\ref{HvsL} of a finitary process is {\em
phenomenological}, in the sense that $H(L)$ and the other quantities
indicated derive only from the observed distribution $\Pr (s^L)$ over 
sequences. In particular, they do not require any additional or prior
knowledge of the source and its internal structure.

\subsection{Discrete Derivatives and Integrals}

We begin by briefly collecting some elementary properties of discrete
derivatives.  Consider an arbitrary function $F: \mathbb{Z} 
\rightarrow \mathbb{R}$.  In what follows, the function $F$ will be
the Shannon block entropy $H(L)$, but for now we consider general
functions.  The discrete derivative is the linear operator defined
by:  
\begin{equation}
 (\Delta F)(L) \, \equiv \, F(L) - F(L\!-\!1) \;.
\end{equation}
The picture is that the operator $\Delta$ acts on $F$ to produce a new
function $\Delta F$ which, when evaluated at $L$, yields $F(L) -
F(L\!-\!1)$.  Higher-order derivatives are defined by composition:  
\begin{equation}
  \Delta^n F \, \equiv \,  (\Delta \circ \Delta^{n-1}) F \;,
\end{equation}
where $\Delta^0 F \equiv F$ and $n \geq 1$.
For example, the second discrete derivative
is given by:
\begin{eqnarray}
  \Delta^2 F(L) & \, \equiv \, & (\Delta \circ \Delta)F(L) \\
   & \,  = \, & F(L) - 2F(L\!-\!1) + F(L\!-\!2) \;.
\label{Curvature}
\end{eqnarray}

One ``integrates'' a discrete function $\Delta F(L)$ by summing:
\begin{equation}
\sum_{L=A}^{B} \Delta F(L) \, =\,  F(B) - F(A\!-\!1) \;. 
\label{Integration}
\end{equation}
An integration-by-parts formula also holds:
\begin{eqnarray}
\nonumber
\sum_{L=A}^{B} & L & \Delta F(L) \, =  \,B F(B) \, - \\ 
    & & A F(A\!-\!1) \, -  \sum_{L=A}^{B-1} F(L) \;. 
\label{IntegrationByParts}
\end{eqnarray}
Note the shift in the sum's limits on the right-hand side.

\subsection{$\Delta H(L)$:  Entropy Gain} 

\begin{figure}[tbp]
\epsfxsize=2.7in
\begin{center}
\leavevmode
\epsffile{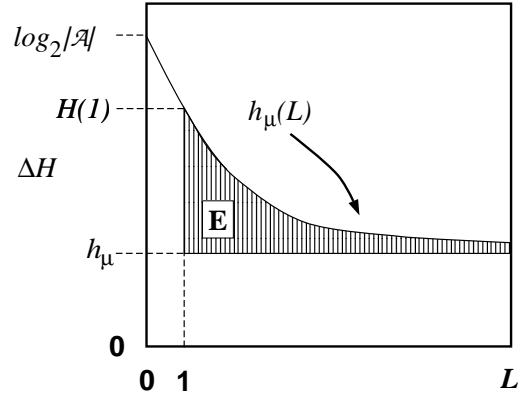}
\end{center}
\vspace{2mm}
\caption{Entropy-rate convergence: A schematic plot of $\hmu(L) =
\Delta H(L)$ versus $L$ using the finitary process's $H(L)$ shown in
Fig.~\ref{HvsL}. The entropy rate asymptote $h_\mu$ is indicated by the
lower horizontal dashed line. The shaded area is the excess entropy
$\bf E$.}
\label{hvsL}
\end{figure}

We now consider the effects of applying the discrete derivative
operator $\Delta$ to the entropy growth curve $H(L)$.  We begin with
the first derivative of $H(L)$: 
\begin{equation}
\Delta H(L) \, \equiv \, H(L) - H(L-1) \;,
\label{EntropyGain}
\end{equation}
where $L > 0$.  The units of $\Delta H(L)$ are {\em bits/symbol}. A
plot of a typical $\Delta H(L)$ vs.~$L$ is shown in Fig.~\ref{hvsL}.
We refer to $\Delta H(L)$ as the {\em entropy gain} for obvious
reasons. 

If a measurement has not yet been made, the apparent entropy rate is
maximal. Thus, we define $\Delta H(0) = \l2 | \Abet | $. In a Bayesian 
modeling setting this is equivalent to being told only that the source
has $|\Abet|$ symbols and then assuming the process is independent
identically distributed and uniformly distributed over individual
symbols.

Having made a single measurement in each experiment in an
ensemble or, equivalently, only looking at single-symbol statistics in
one experiment, the entropy gain is the single-symbol Shannon entropy:
$\Delta H(1) = H(1) - H(0) = H(1)$, since we defined $H(0) = 0$. 

Let's now look at some properties of $\Delta H(L)$.  

\begin{proposition} $\Delta H(L)$ is an information gain: 
\begin{equation}
  \Delta H(L) = {\cal D}[ {\rm Pr}(s^L) || {\rm Pr}(s^{L-1}) ] \;,
\label{1st.derivative}
\end{equation}
\label{DeltaHisInfoGain}
\end{proposition}
where $L > 1$.

\noindent {\em Proof}: Since many of the proofs are straightforward,
direct calculations, we have put most of them in
Appendix \ref{PropProofs} so as not to interrupt the flow of ideas in
the main sections.  Proposition \ref{DeltaHisInfoGain} is proved in
App.~\ref{ProofDeltaHisInfoGain}.$\Box$ 

Note that Eq.~(\ref{1st.derivative}) is a slightly different form for
the information gain than that defined in Eq.~(\ref{info.gain.def}).
Unlike Eq.~(\ref{info.gain.def}), in Eq.~(\ref{1st.derivative}) the
two distributions do not have the same support:  one $\{ s^L \}$ is a
refinement of the other $\{ s^{L-1} \}$. When this is the case, we
extend the length $L-1$ distribution to a distribution over length $L$
sequences by concatenating the symbols $s_{L-1}$ with equal probability
onto $s_0, \ldots, s_{L-2}$. We then sum the terms in ${\cal D}$ over
the set of length $L$ sequences.

Note that since the information gain is a non-negative quantity
\cite{Cove91}, it follows from Prop.~\ref{DeltaHisInfoGain} that
$\Delta H(L) \equiv H(L) - H(L\!-\!1) \geq 0$, as remarked earlier.  
In a subsequent section, we shall see that $\Delta^2H(L) \leq 0$;
hence, $\Delta H(L)$ is monotone decreasing.

The derivative $\Delta H(L)$ may also be written as a conditional
entropy. Since
\begin{equation}
\frac{ {\rm Pr} (s^L) } { {\rm Pr} (s^{L-1}) }
  = {\rm Pr}(s_L|s^{L-1}) ~.
\end{equation}
it immediately follows from Eq.~(\ref{EntropyGain}) that
\begin{equation}
\label{DeltaHasCondEnt}
  \Delta H(L) \, = \, H[S_L|S^{L-1}] \;.
\end{equation}
This observation helps strengthen our interpretation of $h_\mu$. 
Recall that the entropy rate $h_\mu$ was defined in
Eq.~(\ref{ent.def}) as $\lim_{L\rightarrow \infty} H(L)/L$.  As is
well known (see, e.g., Ref.~\cite{Cove91}), the entropy rate may also
be written as: 
\begin{equation}
  h_\mu \, = \, \lim_{L \rightarrow \infty} H[S_L|S^{L-1}] \;.
\end{equation}
That is, $h_\mu$ is the average uncertainty of the variable $S_L$,
given that an arbitrarily large number of preceding symbols have been
seen.  

By virtue of Eq.~(\ref{DeltaHasCondEnt}), we see that
\begin{equation}
\hmu = \lim_{L \rightarrow \infty} \Delta H(L) \;.
\label{EntropyRate}
\end{equation}
Following Refs.~\cite{Crut82b,Crut82c,Pack82,Crut97a,Feld98a},
we denote $\Delta H(L)$ by $h_\mu(L)$:
\begin{eqnarray}
  h_\mu(L) & \equiv & \Delta H(L) \\ \nonumber
  & \equiv & H(L) - H(L-1) \;, ~L \geq 1 ~.
\end{eqnarray}

The function $\hmu (L)$ is the estimate of how random the source
appears if only blocks of variables up to length $L$ are considered.
Thus, $h_\mu(L)$ may be thought of as a finite-$L$ approximation to
the entropy rate $h_\mu$ --- the {\em apparent entropy rate} at length
$L$. Alternatively, the entropy rate $\hmu$ can be estimated for
finite $L$ by appealing to its original definition \cite{Shan62}, i.e.,
Eq.~(\ref{ent.def}). We thus define another finite-$L$ entropy rate
estimate:
\begin{equation}
  h_\mu^\prime (L) \equiv \frac{H(L)}{L} \;, ~L \geq 1 ~,
\end{equation}
where we also take $h_\mu^\prime (0) \equiv \l2 \Abet$. Note that
while we have 
\begin{equation}
  \lim_{L \rightarrow \infty} h_\mu^\prime (L)
  = \lim_{L \rightarrow \infty} h_\mu (L) \;,
\end{equation}
in general, it is the case that
\begin{equation}
 h_\mu^\prime (L) \, \neq \, h_\mu (L) \;, ~L < \infty \;.
\end{equation}
Moreover, $\hmu^\prime (L)$ converges more slowly than $\hmu (L)$.
(The examples later illustrate the slow convergence.)

\begin{lemma}:
\begin{equation}
  h_\mu^\prime (L) \geq h_\mu (L) \geq h_\mu \;.
\end{equation}
\label{EntropyRateConvergeFromAbove}
\end{lemma}

\noindent {\em Proof}: See
App. \ref{ProofEntropyRateConvergeFromAbove}. $\Box$

\subsection{Entropy Gain and Redundancy}

The entropy gain can also be interpreted as a type of redundancy. To
see this, first recall that the redundancy, Eq.~(\ref{Redundancy}),
is the difference between $\log_2 | \Abet |$ and $h_\mu$, where
$\log_2 | \Abet |$ is the entropy given no knowledge of the source
apart from the alphabet size, and $h_\mu$ is the entropy of the source
given knowledge of the distribution of arbitrarily large $L$-blocks.
But what is the redundancy if the observer already knows the actual
distribution $\Pr (s^L)$ of words up to length $L$?  

This question is answered by the $L$-redundancy:
\begin{equation}
{\bf R}(L) \, \equiv \,  H(L) - \hmu L \;.
\label{IntrinsicRedundancy}
\end{equation}
Here, $H(L)$ is the entropy given that $\Pr (s^L)$ is known, and 
the product $\hmu L$ is the entropy of an $L$-block if one uses
only the asymptotic form of $H(L)$ given in
Eq.~(\ref{SimpleScalingForm}).  Note that ${\bf R}(L) \leq {\bf R}$,
where ${\bf R}$ is defined in Eq.~(\ref{Redundancy})

We now define the per-symbol $L$-redundancy:
\begin{equation}
{\bf r} (L)\, \equiv \, \Delta {\bf R}(L) \, \equiv \, \hmu(L) - \hmu \;.
\label{RofL.def}
\end{equation}
The quantity ${\bf r}(L)$ gives the difference between the per-symbol
entropy conditioned on $L$ measurements and the per-symbol entropy
conditioned on an infinite number of measurements.  In other words, 
${\bf r} (L)$ measures the extent to which the length-$L$ entropy rate
estimate exceeds the actual per-symbol entropy. Any difference
indicates that there is redundant information in the $L$-blocks in the 
amount of ${\bf r} (L)$ bits.  Ebeling \cite{Ebel97b} refers to $r(L)$
as the local (i.e., $L$-dependent) predictability.

\subsection{$\Delta^2 H(L)$: Predictability Gain}

\begin{figure}[tbp]
\epsfxsize=2.4in
\begin{center}
\leavevmode
\epsffile{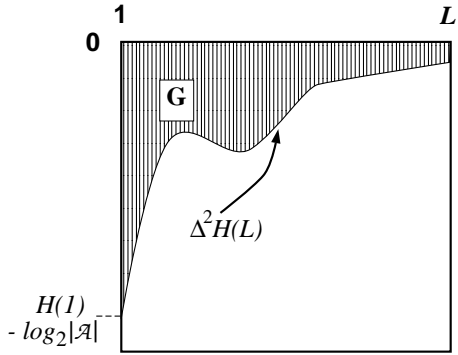}
\end{center}
\vspace{2mm}
\caption{Predictability gain: A schematic plot of $\Delta^2 H(L)$
versus $L$ using the ``typical'' $\hmu(L)$ shown in
Fig.~\ref{hvsL}. The shaded area is the total predictability $\TP$.} 
\label{gvsL}
\end{figure}

If we interpret $\hmu (L)$ as an estimate of the source's
unpredictability and recall that it decreases monotonically to $\hmu$, 
we can look at $\Delta^2 H(L)$ --- the rate of change of $\hmu (L)$
--- as the rate at which unpredictability is lost. Equivalently, we
can view $-\Delta^2 H(L)$ as the improvement in our predictions in
going from $L-1$ to $L$ blocks.  This is the change in the entropy
rate estimate $h_\mu(L)$ and is given by the {\em predictability
gain}: 
\begin{equation}
\Delta^2 H(L) \equiv \Delta h_\mu(L) \,=\, \hmu(L) - \hmu(L\!-\!1) \;, 
\end{equation}
where $L > 0$; the units of $\Delta^2 H(L)$ are {\em bits/symbol$^{\,2}$}.
(See Fig. \ref{gvsL}.)
Since we defined $h_\mu(0) \equiv \l2 | \Abet|$, we have that
\begin{equation}
\Delta^2 H(1) = H(1) - \l2 | \Abet | ~.
\end{equation}
The quantity $\Delta^2 H(0)$ is not defined. 

A large value of $|\Delta^2 H(L) |$ indicates that going from
statistics over $(L\!-\!1)$-blocks to $L$-blocks reduces the
uncertainty by a large amount.  Speaking loosely, we shall see in
Sec.~\ref{Examples_Section} that a large value of $|\Delta^2 H(L) |$
suggests that the $L^{\rm th}$ measurement is particularly
informative.  

\begin{proposition} $\Delta^2 H(L)$ is a conditional information gain:
\begin{equation}
  \Delta^2 H(L) = - {\cal D}
  [ {\rm Pr}(s_{L-1}|s^{L-2}) || {\rm Pr}(s_{L-2}|s^{L-3}) ] \;,
\label{2nd.derivative}
\end{equation}
for $L >3$. 
\label{DeltaSqrHisCondlInfoGain}
\end{proposition}

\noindent {\em Proof:} See App.~\ref{ProofDeltaSqrHisCondlInfoGain}.
$\Box$

\noindent Since the information gain is non-negative, it follows from
Prop.~\ref{DeltaSqrHisCondlInfoGain} that $\Delta^2 H(L) \leq 0$ and
so $H(L)$ is a concave function of $L$. 

The observation contained in Prop.~\ref{DeltaSqrHisCondlInfoGain}
first appeared in Refs.~\cite{Erik87} and \cite{Lind89a}. There,
$-\Delta^2 H(L)$ is referred to as the {\em correlation
information}\/.  However, we feel that the term ``predictability
gain'' is a more accurate name for this quantity. The quantity
$-\Delta^2 H(L)$ measures the reduction in per-symbol uncertainty in
going from $(L\!-\!1)$- to $L$-block statistics. While $-\Delta^2
H(L)$ is related to the correlation between symbols $L$ time steps
apart, it does not directly measure their correlation.  The
information theoretic analog of the two-variable correlation function
is the mutual information between symbols $L$ steps apart:
$I[S_t;S_{t+1}]$, averaged over $t$.  For a discussion of two-symbol
mutual information and how they compare with correlation functions,
see Refs.~\cite{Li90} and \cite{Herz95}.

\subsection{Entropy-Derivative Limits}
\label{ent.limit.section}

Ultimately, we are interested in how $H(L)$ and its derivatives
converge to their asymptotic values. As we will now show, this
question is well posed because the derivatives of $H(L)$ have well
defined limiting behavior. First, as mentioned above, for stationary
sources, $\lim_{L \rightarrow \infty} \Delta H(L) \, = \, \hmu$.  An
immediate consequence of this is is the following.

\begin{lemma}
For stationary processes, the higher derivatives of $H(L)$ vanish in
the $L \rightarrow \infty$ limit:  
\begin{equation}
  \lim_{L \rightarrow \infty} \Delta^n H(L) = 0, \; n \geq 2 \;.
\label{ent.derivative.vanish}
\end{equation}
\end{lemma}

\noindent {\em Proof.} To see this, first recall that the limit
$h_\mu = \lim_{L \rightarrow \infty} \Delta H(L)$ exists for a
stationary source \cite{Cove91} and so the sequence
$\Delta H(0), \Delta H(1), \Delta H(2), \ldots$ converges. It follows
from this that $\lim_{L \rightarrow \infty} [ \Delta H(L) - \Delta
H(L-1)] = \lim_{L \rightarrow \infty}  \Delta^2 H(L) = 0$.  This
proves the $n=2$ case of Eq.~(\ref{ent.derivative.vanish}).  The
$n\geq 3$ cases of Eq.~(\ref{ent.derivative.vanish}) then follow via
identical arguments.$\Box$   

To recapitulate, for the finitary processes we are considering in the
$L \rightarrow \infty$ limit we have that
\begin{equation}
H(L) \sim h_\mu L ~,
\end{equation}
plus possible sublinear terms.
We also have that
\begin{equation}
  \lim_{L \rightarrow \infty}  \Delta H(L)  \, = \, h_\mu \;,
\end{equation}
and
\begin{equation}
  \lim_{L \rightarrow \infty}  \Delta^n H(L)  \, = 0 \;, \;\; {\rm for}
  \; n \geq 2\;.  
\end{equation}

\section{Entropy Convergence Integrals}
\label{Convergence_Integrals}
\label{EntropyIntegrals}

Since limits at each level of the entropy-derivative hierarchy exist,
we can ask {\em how} the derivatives converge to their limits by
investigating the following ``integrals'': 
\begin{equation}
 {\cal I}_n \equiv \sum_{L=L_{\rm n}}^\infty
 \left[ \Delta^n H(L)
 - \lim_{L \rightarrow \infty} \Delta^n H(L) \right] \;.
\label{General.Integral.def}
\end{equation}
The lower limit $L_{\rm n}$ is taken to be the first value of $L$
at which $\Delta^n H(L)$ is defined.  The picture here is that
at each $L$, $\Delta^n H(L)$ over- or under-estimates the
asymptotic value $\lim_{L \rightarrow \infty} \Delta^n H(L)$ by an
amount $\Delta^n H(L) - \lim_{L \rightarrow \infty} \Delta^n H(L)$.
Summing up all of these estimates provides a measure, perhaps
somewhat coarse, of the manner in which an entropy derivative
converges to its asymptotic value. The larger the sum, the slower the
convergence.  The latter, in turn, indicates correlations within
larger $L$ sequences, thus suggesting that a process possesses a
greater degree of structure --- internal structure that is responsible for
maintaining the correlations.  

\subsection{Predictability}

We first examine ${\cal I}_2$.  Recall that
$\lim_{L \rightarrow \infty} \Delta^2 H(L) = 0$ and that
$\Delta^2 H(L)$ is defined for $L \geq 1$.  For reasons that will
become clear shortly, we refer to ${\cal I}_2$ as the 
{\em total predictability} $\TP$. It is defined as:
\begin{equation}
{\TP} \, \equiv \, {\cal I}_2 \, = \sum_{L=1}^\infty \Delta^2
H(L) \;, 
\label{TotalPredictability}
\end{equation}
Geometrically, $\TP$ is the area under the $\Delta^2 H(L)$ curve, as
shown in Fig.~\ref{gvsL}. The units of $\TP$ are {\em bits/symbol}, as
may be inferred geometrically from Fig.~\ref{gvsL}, where the units of
the horizontal axis are bits and those of the vertical axis are
bits/symbol$^{\,2}$.   Alternatively, this
observation follows directly from
Eq.~(\ref{total.predictability.as.average.corr.info}), when one takes
into account the implied $\Delta L$ $(=1$) in the sum.  An
interpretation of $\TP$ is established by the following result.  

\begin{proposition} \cite{Erik87,Lind89a} The magnitude of the total
predictability is equal to the redundancy, 
  Eq.~(\ref{Redundancy}): 
\begin{equation}
{\TP} = - {\bf R}\;.
\label{total.predictability.is.redundancy}
\end{equation}
\label{TotalPredisRedundancy}
\end{proposition}

\noindent {\em Proof:} See App.~\ref{ProofTotalPredisRedundancy}. $\Box$

This establishes an accounting of the maximum possible information
$\l2 | \Abet |$ available from the measurement channel in terms of
intrinsic randomness $\hmu$ and total predictability $\TP$:
\begin{equation}
\l2 | \Abet | = |{\TP}| + \hmu.
\label{TP.and.hmu}
\end{equation}
That is, the raw information $\l2 | \Abet |$ obtained when making a
single-symbol measurement can be considered to consist of two kinds of 
information: that due to randomness $h_\mu$, on the one hand, and
that due to order or redundancy in the process $\TP$, on the other
hand. 

Alternatively, we see that $\TP = \l2 |\Abet| - h_\mu$.  Thus, viewing
$h_\mu$ as measuring the unpredictable component of a process, and
recalling that $\l2 |\Abet|$ is the maximum possible entropy per
symbol, it follows that $\TP$ measures is the source's predictable
component.  For this reason we refer to $\TP$ as the total
predictability.  Note that this result turns on {\em defining} the
appropriate boundary condition as $\hmu (0) = \l2 | \Abet |$.

There is another form for $\TP$ that provides an additional
interpretation.  The total predictability can be expressed as an
average number of measurement symbols, or average length, where
the average is weighted by the third derivative, $\Delta^3 H(L)$. 

\begin{proposition}
The total predictability can be expressed as a type of average length,
where the average is weighted by the third derivative, $\Delta^3
H(L)$. 
\begin{equation}
\TP = - \sum_{L=2}^\infty (L-1) \Delta^3 H(L) \;,
\label{total.predictability.as.average.corr.info}
\end{equation}
when the sum is finite.
\label{TotalPredisCorrInfo}
\end{proposition}

\noindent {\em Proof:} See App.~\ref{ProofTotalPredisCorrInfo}. $\Box$

Eq.~(\ref{total.predictability.as.average.corr.info}) shows that if
$\Delta^3 H(L)$ is slow to converge to $0$, then $\TP$ will be
large.  Ignoring dimensional considerations, $\TP$ can be viewed as
an average length, since
Eq.~(\ref{total.predictability.as.average.corr.info}) expresses $\TP$ 
as $L$ averaged by $\Delta^3H(L)$.  (Note, however, that $\TP$ is not a
correlation length; a correlation length is typically defined as the
$L$ at which a correlation function has decayed to $1/e$ of its
maximum.)  Alternatively, $\TP$ can be viewed as an average of
$\Delta^3 H(L)$, weighted by $L$.  

Speaking informally, $\TP$ could be viewed as a measure of
``disequilibrium'', since it measures the difference between the
actual entropy rate $\hmu$ and the maximum possible entropy rate $\l2
| \Abet |$. The extent to which $\hmu$ falls below the maximum
measures the deviation from uniform probability, which some authors
have interpreted as an equilibrium condition.  In this vein, several
have proposed complexity measures based on multiplying $\TP$ by
$h_\mu$ \cite{Lope95,Shin99}. However, we and others have shown that
this type of complexity measure fails to capture structure or memory,
since they are only a function of disorder $\hmu$
\cite{Feld98b,Crut00a,Bind00}.  For additional critiques of this type
of complexity measure, see Refs.~\cite{Ebel99a,Andr00}.  

Finally, note that for any periodic process, $\TP = \l2 |\Abet|$,
since $\hmu = 0$.  The total predictability assumes its maximum value
for a completely predictable process.  However, $\TP$ does not tell us
how difficult it is to carry out this prediction, nor how many symbols
must be observed before the process can be optimally predicted.  To
capture these properties of the system, we need to look at other
entropy convergence integrals.

\subsection{Excess Entropy}

Having looked at how $\Delta^2 H(L)$ converges to $0$, we now ask: How
does $\Delta H(L) = h_\mu(L)$ converge to $h_\mu$?  One answer to
this question is provided by ${\cal I}_1$.  For reasons that will be
discussed below, we refer to  ${\cal I}_1$ as the {\em excess entropy}
$\EE$:
\begin{equation}
\EE \, \equiv \, {\cal I}_1 \, = \, \sum_{L=1}^\infty [\hmu(L) -
\hmu] \;,
\label{E.def}
\end{equation}
The units of $\EE$ are {\em bits}. We may view $\EE$ graphically as the
area indicated in the entropy-rate convergence plot of Fig.~\ref{hvsL}.
For now, let us assume that the above sum is finite.  For many cases
of interest, however, this assumption turns out to not be correct; a
point to which we shall return at the end of this section.   

The excess entropy has a number of different interpretations, which
will be discussed below. Excess entropy also goes by a variety of
different names.  References~\cite{Crut83a,Crut97a,Feld98b} use
the term ``excess entropy''. Reference~\cite{Shaw84} uses ``stored
information'' and Refs.~\cite{Gras86,Lind88b,Lind89a,Erik87,Ebel99a}
use ``effective measure complexity''. References~\cite{Li91,Arno96}
refer to the excess entropy simply as ``complexity''. References
\cite{Bial00a,Neme00a} refer to the excess entropy as ``predictive
information''. In Refs.~\cite{Csor89a,Kauf91a}, the excess entropy is
called the ``reduced R\'enyi entropy of order $1$''.

\subsubsection{$\EE$ as predictability-gain-weighted length}

\begin{proposition}
The excess entropy may also be written as:
\begin{equation}
\EE = - \sum_{L=2}^\infty (L-1) \Delta^2 H(L) \;.
\label{ExcessEntropy}
\end{equation}
\label{TotalExcessEntropy}
\end{proposition}

\noindent {\em Proof:} See App.~\ref{ProofTotalExcessEntropy}. $\Box$

Eq.~(\ref{ExcessEntropy}) shows that $\EE$ may also be viewed as an
average $L$, weighted by the predictability gain $\Delta^2 H(L)$, a
view emphasized in Ref.~\cite{Lind88b}.  However, this is not a
dimensionally consistent interpretation, since $\EE$ has units of
bits.  Alternatively, Eq.~(\ref{ExcessEntropy}) shows that the excess
entropy can be seen as an average of $\Delta^2 H(L)$, weighted by the
block-length $L$.

\subsubsection{$\EE$ as intrinsic redundancy}

The length-$L$ approximation $h_\mu (L)$ typically overestimates the
entropy rate $h_\mu$ at finite $L$. Specifically, $h_\mu(L)$
overestimates the latter by an amount $h_\mu (L) - h_\mu$ that
measures how much more random single measurements appear knowing
the finite $L$-block statistics than knowing the statistics of infinite
sequences. In other words,
this excess randomness tells us how much additional information must
be gained about the sequences in order to reveal the actual per-symbol
uncertainty $h_\mu$. This merely restates the fact that the difference
$h_\mu (L) - h_\mu$ is the per-symbol redundancy ${\bf r}(L)$, defined
originally in Eq.~(\ref{RofL.def}).  Though the source
appears more random at length $L$ by the amount ${\bf r}(L)$, this
amount is also the information-carrying capacity in the $L$-blocks
that is not actually random, but is due instead to correlations. We
conclude that entropy-rate convergence is controlled by this redundancy
in the source. Presumably, this redundancy is related to structures
and memory intrinsic to the process.  However, specifying 
how this memory is organized cannot be done within the framework of
information theory; a more structural approach based on the theory of
computation must be used. We return to the latter in the conclusion.

There are many ways in which the finite-$L$ approximations
$h_\mu (L)$ can converge to their asymptotic value $h_\mu$. (Recall
Fig.~\ref{hvsL}.) Fixing the values of $H(1)$ and $h_\mu$, for
example, does not determine the form of the $h_\mu(L)$ curve. At each
$L$ we obtain additional information about how $h_\mu(L)$ converges,
information not contained in the values of $H(L)$ and $h_\mu(L)$ at
smaller $L$. Thus, roughly speaking, each $h_\mu(L)$ is an independent
indicator of the manner by which $h_\mu(L)$ converges to $h_\mu$.  

Since each increment $h_\mu(L) - h_\mu$ is an independent contribution
in the sense just described, one sums up the individual per-symbol
$L$-redundancies to obtain the total amount of apparent memory in a
source \cite{Shaw84,Gras86,Crut83a,Szep86,Lind88b,Li91,Lind88a,Junc79}.
Calling this sum {\em intrinsic redundancy}, we have the following
result.

\begin{proposition} The excess entropy is the intrinsic redundancy of
the source: 
\begin{equation}
  \EE = \sum_{L=1}^\infty {\bf r} (L) \;.
\end{equation}
\label{EEfromRedundancy}
\end{proposition}

\noindent {\em Proof}: This follows directly from inserting the
definition of intrinsic redundancy, Eq.~(\ref{RofL.def}), in
Eq.~(\ref{E.def}). $\Box$ 

The next proposition establishes a geometric interpretation of $\EE$ and
an asymptotic form for $H(L)$.

\begin{proposition} The excess entropy is the subextensive part of
$H(L)$: 
\begin{equation}
  \EE = \lim_{L \rightarrow \infty} [ H(L) - \hmu L ]\;.
\end{equation}
\label{EEfromEntropyGrowth}
\end{proposition}

\noindent {\em Proof}: See App.~\ref{ProofEEfromEntropyGrowth}. $\Box$

This proposition implies the following asymptotic form for $H(L)$: 
\begin{equation}
  H(L) \, \sim \, \EE + \hmu L \,, {\rm as} \; L \rightarrow \infty \;.
\label{EEScalingForm}
\end{equation}
Thus, we see that $\EE$ is the $L = 0$ intercept of the linear
function Eq.~(\ref{EEScalingForm}) to which $H(L)$ asymptotes. This
observation, also made in
Refs.~\cite{Bial00a,Neme00a,Shaw84,Gras86,Li91}, is shown graphically
in Fig.~\ref{HvsL}.  Note that $\EE \geq 0$, since $H(L) \geq \hmu L$.
Note also that if $h_\mu = 0$, then $\EE = \lim_{L \rightarrow \infty}
H(L)$.  

A useful consequence of Prop.~\ref{EEfromEntropyGrowth} is that it
leads one to use Eq.~(\ref{EEScalingForm}) instead of the original
(very simple) scaling of Eq.~(\ref{SimpleScalingForm}). Later sections 
address how ignoring Eq.~(\ref{EEScalingForm}) leads to erroneous
conclusions about a process's unpredictability and structure.

\subsubsection{$\EE$ as mutual information}

Yet another way to understand excess entropy is through its expression
as a mutual information. 

\begin{proposition} The excess entropy is the mutual information
between the left and right (past and future) semi-infinite halves of
the chain $\all$:    
\label{EandMI_Proposition}
\begin{equation}
  \EE \, = \,\lim_{L \rightarrow \infty}  I[S_0 S_1 \cdots S_{2L-1};
  S_{2L} S_{2L+1} S_{2L-1}] \;. 
\label{EandMI}
\end{equation}
when the limit exists.
\end{proposition}

\noindent {\em Proof}: See App.~\ref{ProofEandMI}. $\Box$

Note that $\EE$ is not a two-symbol mutual information, but is instead
the mutual information between two semi-infinite blocks of variables.

Eq.~(\ref{EandMI}) says that $\EE$ measures the amount of historical
information stored in the present that is communicated to the future.
For a discussion of some of the subtleties associated with this
interpretation, however, see Ref.~\cite{Shal98a}.

Prop.~\ref{EandMI_Proposition} also shows that $\EE$ can be
interpreted as the {\em cost of amnesia}: If one suddenly loses track
of a source, so that it cannot be predicted at an error level
determined by the entropy rate $\hmu$, then the entire string appears
more random by a total of $\EE$ bits.

\subsubsection{Finitary processes}
\label{finitary.section}

We have argued above that the excess entropy $\EE$ provides a measure
of one kind of memory. Thus, we refer to those processes
with a finite excess entropy as finite-memory sources or, simply,
{\em finitary processes}, and those with infinite memory, {\em
infinitary processes}. 

\begin{definition}
A process is {\em finitary} if its excess entropy is finite.  
\end{definition}

\begin{definition}
A process is {\em infinitary}\/ if its excess entropy is infinite.  
\end{definition}

\begin{proposition} For finitary processes the entropy-rate estimate
$\hmu(L)$ decays faster than $1/L$ to the entropy rate $\hmu$. That is,
\begin{equation}
\hmu(L) - \hmu < \frac{A}{L} \;,
\label{FinitaryEntropyRateDecayForm}
\end{equation}
for large $L$ and where $A$ is a constant. For infinitary processes
$\hmu(L)$ decays at or slower than $1/L$.
\label{FinitaryEntropyRateDecay}
\end{proposition}

\noindent {\em Proof}: By direct inspection of Eq.~(\ref{E.def}). $\Box$

One consequence is that the entropy growth for finitary processes scales
as $H(L) \sim \EE + h_\mu L$ in the $L \rightarrow \infty$ limit, where
$\EE$ is a constant, independent of $L$.  In contrast, an infinitary
process might scale as 
\begin{equation}
H(L) \, \sim \, c_1 + c_2\log L + h_\mu L ~,
\end{equation}
where $c_1$ and $c_2$ are constants.  For such a system, the excess
entropy $\EE$ diverges logarithmically and $h_\mu(L) - h_\mu \sim
L^{-1}$. 

In Sec.~\ref{Examples_Section} we shall determine $\EE$, $h_\mu$, and
related quantities for several finitary sources and one infinitary source.
There are, however, a few particularly simple classes of finitary
process for which one can obtain general expressions for $\EE$, which
we state here before continuing.

\begin{proposition} For a periodic process of period $p$, the excess
entropy is given by
\begin{equation}
  \EE = \l2 \; p ~.
\label{EEPeriodic}
\end{equation}
\end{proposition}

\noindent {\em Proof}: One observes that $H(L) = \l2 p$, for $L >
p$. $\Box$ 

\begin{proposition} For an order-$R$ Markovian process, the excess
entropy is given by
\begin{equation}
  \EE = H(R) - R \hmu \;.
\label{EEMarkovian}
\end{equation}
Recall that an order-$R$ Markovian process was defined in
Eq.~(\ref{OrderR.Markovian.def}). 
\end{proposition}

\noindent {\em Proof}:  This result will be proved in
Sec.~\ref{Markovian_Example}, when we consider an example Markovian
process.  Also, see Refs.~\cite{Gatl72}, \cite{Crut97a}, and
\cite{Feld98a}. $\Box$ 

For finitary processes that are not finite-order Markovian, the entropy-rate
estimate $\hmu(L)$ often decays exponentially to the entropy rate $\hmu$: 
\begin{equation}
\hmu(L) - \hmu \, \sim \, A 2^{- \gamma L} \;,
\label{EntropyRateDecayForm}
\end{equation}
for large $L$ and where $\gamma$ and $A$ are constants.

Exponential decay was first observed for various kinds of one-dimensional
map of the interval and a scaling theory was developed based on that ansatz
\cite{Crut82b}. Later Eq.~(\ref{EntropyRateDecayForm}) was proven to hold
for one-dimensional, fully chaotic maps with a unique invariant ergodic
measure that is absolutely continuous with respect to the Lebesgue measure
\cite{Csor93b}. To our knowledge, there is not a direct proof of exponential
decay for more general finitary processes. There is, however, a large amount
of empirical evidence suggesting this form of convergence
\cite{Gras86,Gyor85,Szep89a,Crut82b}. Nevertheless, several lines of
reasoning suggest that exponential decay is typical and to be expected.
For further discussion, see Appendix \ref{ExponentialDecayAppendix}.  

\begin{corollary} For exponential-decay finitary processes the excess entropy
is given by 
\begin{equation}
\EE \approx \EE_\gamma \equiv \frac{H(1) - \hmu}{1 - 2^{-\gamma}} \;, 
\label{EERelatedToGamma}
\end{equation}
where $\gamma$ is the decay exponent of Eq.~\ref{EntropyRateDecayForm}
and $H(1)$ is the single-symbol entropy.
\end{corollary}

\noindent {\em Proof}: One directly calculates the area between two
curves in the entropy convergence plot of Fig.~\ref{hvsL}. The first
is the constant line at $\hmu$. The second is the curve specified by
Eq.~\ref{EntropyRateDecayForm} with the boundary condition $\hmu(1) =
H(1)$.  Alternatively, Eq. (\ref{EntropyRateDecayForm}) may be inserted
into Eq.~(\ref{ExcessEntropy}); Eq.~(\ref{EERelatedToGamma}) then
follows after a few steps of algebra. $\Box$

Note that Eq.~(\ref{EERelatedToGamma}) is an approximate result; it is 
exact only if Eq.~(\ref{EntropyRateDecayForm}) holds for all $L$.  In
practice, for small $L$ $h_\mu(L) - h_\mu$ is larger than its
asymptotic form $A2^{-\gamma L}$ and, thus, $\EE_\gamma$ gives an upper
bound on $\EE$.   

\subsubsection{Finite-$L$ expressions for $\EE$}
\label{EEConvergence.section}

There are at least two different ways to estimate the excess entropy
$\EE$ for finite $L$. First, we have the partial-sum estimate given by
\begin{eqnarray}
\nonumber
  \EE(L) & \equiv & H(L) - L h_\mu(L) \\
  & = & \sum_{M=1}^L \left[ h_\mu(M) - h_\mu(L) \right] \;.
\label{EEEstimate}
\end{eqnarray}
The second equality follows immediately from the integration formula,
Eq.~(\ref{Integration}), and the boundary condition $H(0) = 0$. 

Alternatively, a finite-$L$ excess entropy can be defined as the
mutual information between $L/2$-blocks:
\begin{equation}
\EE^\prime (L) \, \equiv \, I[S_0 S_1 \cdots S_{L/2 - 1}; S_{L/2}
S_{L/2+1} S_{L-1}] \;,
\label{EEPrimeEstimate}
\end{equation}
for $L$ even.  If $L$ is odd, we define $\EE^\prime (L) =
\EE^\prime (L-1)$.  The expression in Eq.~(\ref{EEPrimeEstimate}),
however, is not as good an estimator of ${\bf E}$ as that of equation
Eq.~(\ref{EEEstimate}), as established by the following lemma:

\begin{lemma}:
\begin{equation}
\EE^\prime (L) \leq \EE (L) \leq \EE \;.
\end{equation}
\label{EEConvergenceBounds}
\end{lemma}

\noindent {\em Proof:} See App. \ref{ProofEEConvergenceBounds}. $\Box$

\section{Transient Information}
\label{T_Section}

Thus far, we have discussed derivatives of the entropy growth curve
$H(L)$, and we have also defined and interpreted two integrals:  the
total predictability $\TP$ and the excess entropy $\EE$. Both $\TP$
and $\EE$ have been introduced previously by a number of authors.  

In this section, however, we introduce a new quantity, by following
the same line of reasoning that led us to the total predictability,
$\TP = {\cal I}_2$, and the excess entropy, $\EE = {\cal I}_1$. That
is, we ask: How does $H(L)$ converge to its asymptote $\EE + \hmu L$?
The answer to this question is provided by ${\cal I}_0$. For reasons
that will become clear below, we shall call $-{\cal I}_0$ the {\em
transient information} $\TI$: 
\begin{equation}
\TI \equiv - {\cal I}_0 =
  \sum_{L=0}^\infty \left[ \EE + \hmu L - H(L) \right] \;. 
\label{T.def}
\end{equation}
Note that the units of $\TI$ are {\em bits $\times$ symbols}.

The following result establishes an interpretation of $\TI$.

\begin{proposition} The transient information may be written as:
\begin{equation}
  \TI \, = \, \sum_{L=1}^\infty L \left[ h_\mu(L) - \hmu
  \right] \;. 
\label{TI.as.integral.by.parts}
\end{equation}
\label{T.theorem}
\end{proposition}

\noindent {\em Proof:} The proof is a straightforward calculation,
however, since it is a new result, we include it here.  We begin by
writing the right-hand side of Eq.~(\ref{TI.as.integral.by.parts}) as
a partial sum: 
\begin{eqnarray}
  \sum_{L=1}^\infty L \left[ h_\mu(L) - \hmu \right] & \, = &
  \nonumber \\
   & & \hspace{-2cm} \lim_{M \rightarrow \infty}  \sum_{L=1}^M [L
  \Delta H(L) - h_\mu L ]  \;. 
\end{eqnarray}
Using Eq.~(\ref{IntegrationByParts}), this becomes:
\begin{eqnarray}
  \sum_{L=1}^\infty L \left[ h_\mu(L) - \hmu \right] & \, = &
  \nonumber \\
   & & \hspace{-3cm} \lim_{M \rightarrow \infty} \left\{ MH(M) - 
	\sum_{L=1}^{M-1}H(L)  - \sum_{L=1}^M h_\mu L ] \right\}  \;. 
\end{eqnarray}
Using $MH(M) = \sum_{L=0}^{M-1}H(M)$ and $\lim_{M \rightarrow \infty}
H(M) = \EE + h_\mu M$, and rearranging slightly, we have:
\begin{eqnarray}
  \sum_{L=1}^\infty L \left[ h_\mu(L) - \hmu \right] & \, = &
  \nonumber \\
   & & \hspace{-3cm} \lim_{M \rightarrow \infty} \left\{  
	\sum_{L=0}^{M-1}[\EE - H(L) ]  + \sum_{L=0}^{M-1} h_\mu M -
  \sum_{L=1}^M h_\mu L  \right\}  \;.
\label{T.proof.step}  
\end{eqnarray}
But,
\begin{eqnarray}
 \sum_{L=0}^{M-1} h_\mu M & - & \sum_{L=1}^M h_\mu L
 \nonumber \\
  & = & h_\mu \left[ M^2 - \frac{1}{2}M(M+1) \right] \\
  & = & h_\mu \frac{1}{2}M(M-1) \\
  & = & \sum_{L=0}^{M-1}h_\mu L \;. 
\end{eqnarray} 
Using this last line in Eq.~(\ref{T.proof.step}), we have
\begin{eqnarray}
  \sum_{L=1}^\infty L \left[ h_\mu(L) - \hmu \right] & \, = &
  \nonumber \\
   & & \hspace{-3cm} \lim_{M \rightarrow \infty} \left\{  
	\sum_{L=0}^{M-1}[\EE +h_\mu L - H(L) ]  \right\}  \;.
\end{eqnarray}
The right-hand side of the above equation is $\TI$, completing the
proof.  $\Box$

Recall that $\EE + \hmu L$ is the entropy growth curve for a finitary
process, as discussed in Sec.~\ref{finitary.section}. Thus, $\TI$ may
be viewed as a sum of redundancies, ($\EE + \hmu L) - H(L)$, between
the source's actual entropy growth $H(L)$ and the $\EE + \hmu L$
finitary-process approximation.

\subsection{$\TI$ and Synchronization Information}
\label{SyncMarkov}

For finitary processes $H(L)$ scales as $\EE + h_\mu L$ for large $L$.
When this scaling form is reached, we say that the observer is
{\em synchronized} to the process. In other words, when
\begin{equation}
\TI (L) \, \equiv \, \EE + \hmu L - H(L) \, = \, 0 \;,
\label{OSSyncCondition}
\end{equation}
we say the observer is synchronized at length-$L$ sequences. 
As we will see below, observer-process synchronization corresponds to
the observer being in a condition of knowledge such that it can predict
the process outputs at an error rate determined by the process's
entropy rate $\hmu$.  

On average, how much information must an observer extract from
measurements so that it is synchronized to the process in the
sense described above? As argued in the previous section, an
answer to this question is given by the transient information $\TI$.  

We now establish a direct relationship between the transient
information $\TI$ and the amount of information required for
observer synchronization to block-Markovian processes.
We begin by stating the question of
observer synchronization information theoretically and fixing
some notation. 

Assume that the observer has a correct model ${\cal M}=\{{\cal V},
T\}$ of a process, where ${\cal V}$ is a set of states and $T$ the
rule governing transitions between states. That is, $T$ is a matrix
whose components $T_{AB}$ give the probability of making a transition
to state $B$, given that the system is in state $A$, where $A,B \in
{\cal V}$. Contrary to the scenario shown in
Fig.~\ref{MeasurementChannel}, in this section we assume that the
observer {\em directly} measures the process's states.  That is, we
have a Markov process, rather than a {\em hidden} Markov process.

The task for the observer is to make observations and determine in
which state $v \in {\cal V}$ the process is.  Once the observer knows
with certainty in which state the process is, the observer is
{\em synchronized} to the source and the average per-symbol
uncertainty is exactly $h_\mu$. We are interested in describing
how difficult it is to synchronize to a directly observed
Markov process.   

The observer's knowledge of ${\cal V}$ is given by a distribution over
the states $v \in {\cal V}$. Let ${\rm Pr}( v | s^L, {\cal M} )$
denote the distribution over $\cal V$ given that the particular
sequence of symbols $s^L$ has been observed. The entropy of this
distribution over the states measures the observer's
average uncertainty in $v \in {\cal V}$:
\begin{equation}
  H[{\rm Pr}( v | s^L, {\cal M}) ] \, \equiv \, -\sum_{v \in {\cal V}}
  {\rm Pr}( v | s^L, {\cal M}) \log_2  {\rm Pr}( v | s^L, {\cal M})
  \;. 
\end{equation}
Averaging this uncertainty over the possible length-$L$ observations,
we obtain the {\em average state-uncertainty}:
\begin{eqnarray}
  {\cal H}(L) \, \equiv & & \nonumber \\
  & & \hspace{-1.5cm} \, - \sum_{s^L} {\rm Pr}( s^L )\sum_{v \in
  {\cal V}} 
  {\rm Pr}( v | s^L, {\cal M}) \log_2  {\rm Pr}( v | s^L, {\cal M}) \;.
\label{script.H.def}
\end{eqnarray}
The quantity ${\cal H}(L)$ can be used as a criterion for
synchronization. The observer is synchronized to the source when
${\cal H}(L) = 0$ --- that is, when the observer is completely
certain about in which state $v \in \cal V$ the mechanism generating
the sequence is.  And thus, when the condition in
Eq. (\ref{OSSyncCondition}) is met, we see that ${\cal H}(L) = 0$, and
the uncertainty associated with the prediction of the model ${\cal
M}$ is exactly $h_\mu$. 

While the observer is still unsynchronized, though, ${\cal H}(L) >
0$.  We refer to the average total uncertainty experienced by an
observer during the synchronization process as the {\em
synchronization information} $\SI$: 
\begin{equation}
\SI \equiv \sum_{L=0}^\infty {\cal H}(L) ~.
\end{equation}
The synchronization information measures the average total information
that must be extracted from measurements so that the observer is
synchronized.   

In the following, we assume that our model is Markovian of order
$R$.  Additionally, we assume that the set of Markovian states ${\cal
V}$ is associated with the $|\Abet^R|$ possible values of $R$
consecutive symbols; henceforth the latter are referred to as {\em
$R$-blocks}.  Specifically, there is a one-to-one correspondence
between the states $v$ and the $R$-blocks, and hence there exists a
one-to-one, invertible function $\varphi: s^R \rightarrow {\cal V}$.
This function $\varphi$ enables us to move back and forth between the
states $v$ and the $R$-blocks.  For example, we may use $\varphi$ to
rewrite the set of states: 
\begin{equation}
  {\cal V} \, = \, \{ \varphi(s_1 s_2 \cdots s_R): s_i \in \Abet, 1
  \leq i \leq R   \} \;. 
\label{state.Rblock.equiv}
\end{equation}

The matrix $T$ gives the transition probabilities between symbol blocks.
Note that the Markovian states are ``sliding'' in the sense that
a transition from one state to another corresponds to a transition
from, say, symbol block $s_0 s_1 \cdots s_{R-1}$ to $s_1 s_2 \cdots
s_R$.  Thus, it is not hard to see that the transition matrix $T$ is
sparse;  there are at most $|\Abet^{R+1}|$ nonzero entries in the
$|\Abet^R \times \Abet^R|$ matrix $T$.  

\begin{theorem}
\label{SyncTheorem}
For a block-Markovian process, the synchronization information $\SI$
is given by:
\begin{equation}
\SI = \TI + \frac{1}{2} R(R+1) \hmu \;.
\label{SyncTheoremEquation}
\end{equation}
\end{theorem}

\noindent {\em Proof:} See App.~\ref{SyncMarkovProof}.$\Box$

Thus, the transient information $\TI$, together with the entropy rate
$h_\mu$ and the order $R$ of the Markov process, measures how
difficult it is to synchronize to a process.  If a system has a large
$\TI$, then, on average, an observer will be highly uncertain about
the internal state of the process while synchronizing to it.  The
transient information measures a structural property of the
system --- a property not captured by the excess entropy $\EE$.  

\begin{corollary} For periodic processes, $\SI = \TI$.
\label{SyncPeriodic}
\end{corollary}

\noindent {\em Proof:}  For periodic processes, $\hmu = 0$.  Plugging
this in to Eq.~(\ref{SyncTheoremEquation}), the corollary
follows. $\Box$ 

In Sec.~\ref{Per5_Example} we shall see that, while the excess
entropy is the same ($\log_2 p$) for all period $p$ processes, the
transient informations are different.  Thus, the transient information
allows one to draw structural distinctions between different periodic
sequences.  

\begin{corollary}
For exponential-decay finitary processes we have
\begin{equation}
  \TI \, \approx \, \TI\gamma \, \equiv \, \frac{H(1) - h_\mu} { ( 1 -
  2^{-\gamma})^2} \;.  
\label{TIRelatedToGamma}
\end{equation}
\end{corollary}

\noindent {\em Proof:}  Inserting Eq.~(\ref{EntropyRateDecayForm})
into the expression for $\TI$ given in
Eq.~(\ref{TI.as.integral.by.parts}), the result follows after several
steps. $\Box$

Combining
Eqs.~(\ref{EERelatedToGamma}) and (\ref{TIRelatedToGamma}), we arrive
at an exact relationship between the approximate expressions for the
excess entropy and the transient information:
\begin{equation}
  \TI_\gamma \, = \, \frac{ \EE_\gamma^2}{ H(1) - h_\mu} \;.
\label{TIRelatedToEE}
\end{equation}

\subsection{Summary}

This completes our exposition of entropy convergence and our method
of differentiating and integrating $H(L)$ to move between levels.
Table \ref{HierarchySynopsis} summarizes the first levels of the
entropy convergence hierarchy as investigated in the preceding sections.

\end{multicols}

\begin{table}
\begin{center}
\begin{tabular}{c c c c c c}
\multicolumn{6}{c}{Entropy-Convergence Hierarchy}\\
\hline
Level & Derivatives & $L_n$ & $L \rightarrow \infty$ Limit &
\multicolumn{2}{ c }{Integrals} \\ 
\hline
    & & & & At Level $n$  & From Level $n+1$ \\
\hline
0 & $H(L)$ & $L_0 = 0$ & $\infty$ or $\l2 p$ &
    $\TI \equiv \sum_{L=0}^\infty [\EE + \hmu L - H(L)]$ &
    $\TI = \sum_{L=1}^\infty (L) [\Delta H(L)-\hmu]$ \\
1 & $\Delta H(L)$ & $L_1 = 0$ & $\hmu$ &
    $\EE \equiv \sum_{L=1}^\infty [\Delta H (L) - \hmu]$ &
    $\EE = - \sum_{L=2}^\infty (L\!-\!1) \Delta^2 H(L)$ \\
2 & $\Delta^2 H(L)$ & $L_2 = 1$ & 0 &
    $\TP \equiv \sum_{L=1}^\infty \Delta^2 H(L)$ &
    $\TP = - \sum_{L=2}^\infty (L\!-\!1) \Delta^3 H(L)$ \\
... & ...& ... & ... & ... & ... \\
n & $\Delta^n H(L)$ & $L_n = n-1$ & 0 &
    $ {\cal I}_n \equiv \sum_{L=L_n}^\infty \left[ \Delta^n H(L)
		  - \lim_{L \rightarrow \infty} \Delta^n H(L) \right] $ &
    $-\sum_{L=L_n}^\infty (L-1) \Delta^{n+1} H(L)$ \\
\end{tabular}
\caption{Moving up and down the first levels of entropy convergence.}     
\label{HierarchySynopsis}
\end{center}
\vspace{.5cm}
\end{table}

\begin{multicols}{2}


\section{Examples}
\label{Examples_Section}

This section analyzes several variously structured processes to
illustrate a range of different entropy convergence behaviors. The
results demonstrate what the preceding quantities --- such as, the
entropy rate, the excess entropy and the transient information --- do
and do not indicate about a process's organization.

\subsection{Independent, Identically Distributed Processes}

\begin{figure}[tbh]
\epsfxsize=3.0in
\begin{center}
\leavevmode
\epsffile{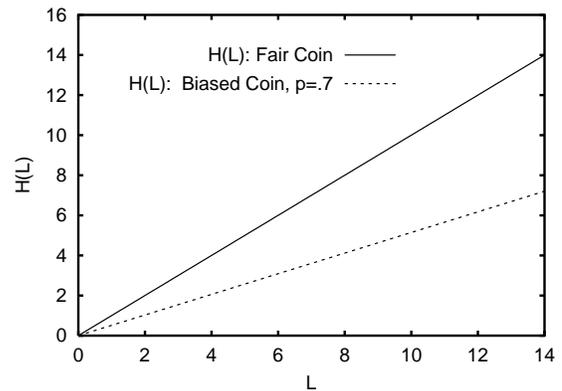}
\end{center}
\vspace{-2mm}
\caption{Entropy growth for IID processes: a fair coin (solid line)
  and a coin (dashed line) with bias $p = 0.7$. }
\label{Coins}
\end{figure}

We begin with the simplest stochastic process: binary variables
independently and identically distributed (IID), as in
Eq.~(\ref{IID}).  Figure \ref{Coins} shows the entropy growth curve
$H(L)$ for two IID processes:  a fair coin and a biased coin with a
bias of $0.7$.   

For both coins $H(L)$ grows linearly. Hence,  $\Delta H(L)$ is
constant for these and all other IID processes.  Note, however, that
the two systems have different entropy rates $h_\mu$.  The fair coin
has an $h_\mu$ of $1$ bit per symbol, while the biased coin, being
less unpredictable, has $h_\mu \approx .8813$.   As a result, from
Eq.~(\ref{TP.and.hmu}) the total predictability $\TP = \log_2 | \Abet
| - h_\mu = 0$ bits for the fair coin and $0.1187$ bits for the
biased coin. The predictability of each process is rather low, as
expected. 

As is clear from Fig.~\ref{Coins}, for both processes the excess
entropy $\EE$ and the transient information $\TI$ are zero.  This
makes sense in light of the interpretations of $\EE$ and $\TI$ given
in the previous sections.  Each coin flip does not depend on past
flips, and so there is no mutual information between the past and the
future. Thus, $\EE = 0$. Similarly, no information is needed to
synchronize to the source --- $H(L)$ assumes its asymptotic form at
$L=1$ --- and so $\TI = 0$. That is, the statistics of isolated flips
are all that is required to optimally predict both processes.
Historical information does not improve predictability.  

\begin{figure}[th]
\begin{center}
\epsfxsize=3.0in
\leavevmode 
\epsffile{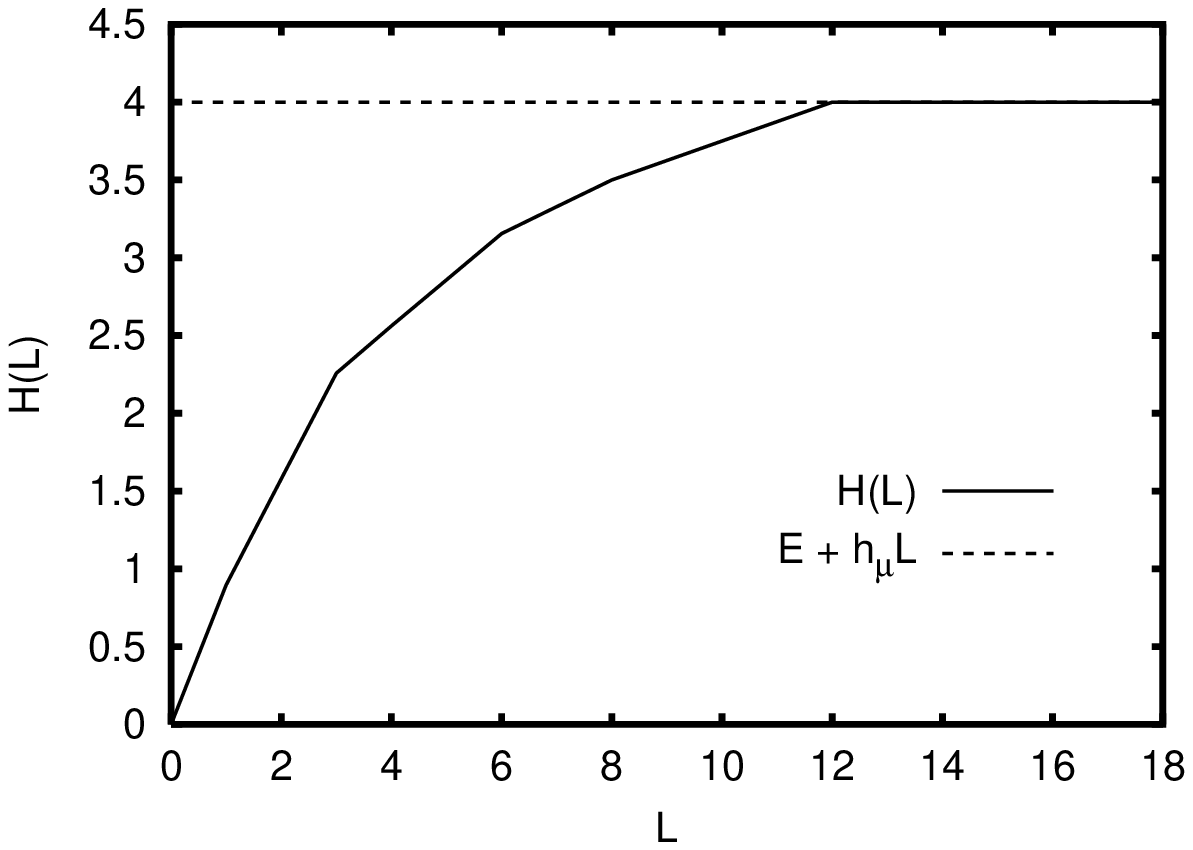} \\
\leavevmode
\mbox{\bf (a)}\\
\epsfxsize=3.0in
\leavevmode
\epsffile{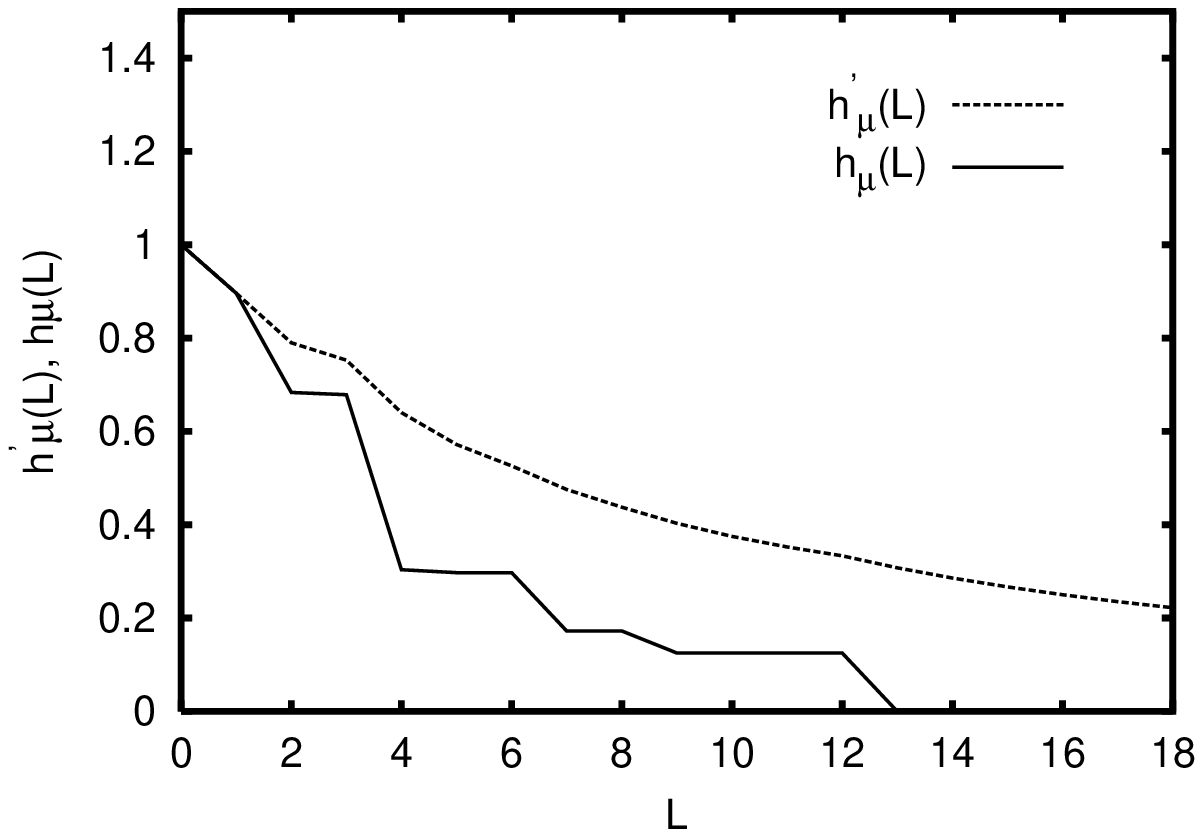} \\
\leavevmode
\mbox{\bf (b)} \\
\end{center}
\caption{Entropy curves for the period-$16$ process:
  $\cdots (1010111011101110)^\infty \cdots$. (a) Entropy growth 
  (solid line) and $\EE + \hmu L$ (dashed line). (b) Entropy
  convergence for the two estimators $\hmu(L)$ (solid line) and
  $\hmu^\prime(L)$ (dotted line).}
\label{Per16}
\end{figure}

\subsection{Periodic Processes}

\subsubsection{A period-$16$ process}

We now consider periodic processes. We begin with a period-$16$
process, whose  $H(L)$ is shown in Fig.~\ref{Per16}(a). The
sequence consists of repetitions of the length-$16$ block $s^{16} =
1010111011101110$.  In Fig.~\ref{Per16}(b) we show the convergence of 
entropy density estimates to the asymptotic value, $h_\mu = 0$.  As
for all period processes, the entropy rate $h_\mu$ for the period-$16$
process is zero; at sufficiently large $L$ the process is perfectly
predictable.  In addition to $h_\mu(L)$, defined above as $H(L) -
H(L-1)$, we show $h^\prime_\mu (L) = H(L) / L$.  The total entropy
converges at $L = 12$. The value of $H(12) = 4$ bits reflects the fact
that there are $16$ equally probable sequences at each $L \geq 12$.   

\begin{figure}[ht]
\begin{center}
\epsfxsize=3.0in
  \leavevmode
  \epsffile{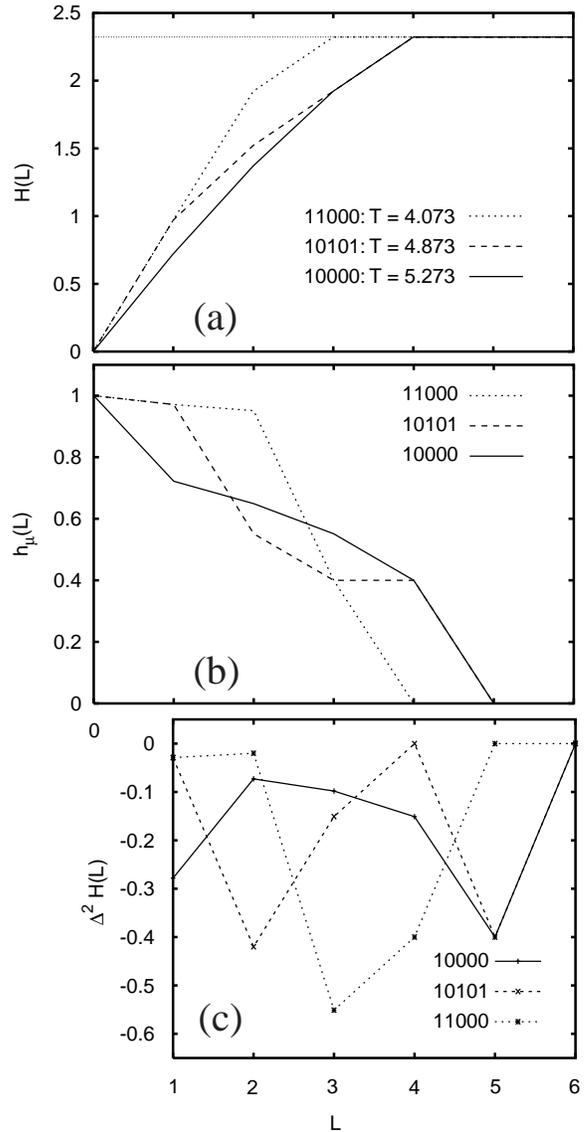}
\end{center}
\vspace{2mm}
\caption{(a) Entropy growth for all period-$5$ processes,
  along with the asymptote $\EE + \hmu L = \log_2 5 \approx 2.321$ 
(thin dashed line). (b) Entropy convergence, for the same period-$5$ 
processes. (c) Predictability gain $\Delta^2 H(L)$.}
\label{Per5}
\end{figure}

\vspace{-0.25in}

\begin{table}[htb]
\begin{center}
\begin{tabular}{c c c}
Template & Number of Observations   & $\TI$  \\
Word     & to Synchronize [symbols] & [bit-symbols] \\
\hline
   11000 &  2.4                   & 4.073 \\
   10000 &  2.8                   & 5.273 \\
   10101 &  3.2                   & 4.873 \\
\end{tabular}
\caption{Synchronizing to period-$5$ processes: Comparing the transient
information $\TI$ to the average number of observations required to
synchronize to the three distinct period-$5$ sequences. }
\end{center}
\label{Per5Sync}
\end{table}

The excess entropy $\EE$ for the period-$16$ process is $\l2 16 = 4$
bits; the sequence's past carries $4$ bits of phase information
about the future. Geometrically, $\EE$ is the vertical-intercept of
the horizontal asymptote on Fig.~\ref{Per16}(a) (dashed line) or the
area under $h_\mu(L)$ on Fig.~\ref{Per16}(b).  The predictability
is $\TP = \l2 2 = 1$ bit per symbol; the system can be predicted
perfectly. 
Finally, the transient information for this period-$16$ process is
$\TI \approx 16.6135$ bit-symbols.  Since this process is Markovian,
Thm. \ref{SyncTheorem} applies.  Thus, we conclude that, on
average, an observer would measure a total uncertainty $\SI$ of
$16.6135$ bits during the process of synchronization. 

\subsubsection{$\TI$ distinguishes period-$p$ processes} 
\label{Per5_Example}

For any periodic process of period $p$, $h_\mu = 0$ and $\EE = \log_2 
p$.  However, there are important structural differences between
different sequences with the same period.  To show this, we consider
all binary period-$5$ processes that are distinct up to permutations
and ($0 \leftrightarrow 1$)-exchanges in their ``template''
words. There are only three such processes:  $(11000)^\infty$,
$(10101)^\infty$, and $(10000)^\infty$.  By the symmetries of the
Shannon entropy function these processes illustrate the only three
types of entropy convergence behavior possible for period-$5$
sequences. 

Figure \ref{Per5}(a) shows the entropy growth curves for each;
Fig.~\ref{Per5}(b) gives the entropy convergence curves; and
Fig.~\ref{Per5}(c) gives the predictability gain $\Delta^2 H(L)$.
By $L = 4$, $H(L)$ converges to $\EE = \log_2 5 \approx 2.321$
bits.  We see that $\hmu(L) = 0$ at this and larger $L$. For all three
processes, $\TP = 1 - h_\mu = 1$ bit per symbol: Again, the
information in each measurement concerns the periodic component of the
process. The predictability gain per measurement vanishes at $L = 6$,
since at that point all length-$5$ templates have been completely
parsed and the process appears completely predictable. It is a useful
exercise in understanding $\Delta^2 H(L)$ to work through each template
symbol-by-symbol to see which symbols are more and less informative
about each template's phase. For example, on the one hand, observing
the fourth symbol of the $(10101)^\infty$ process does not improve
predictability. On the other hand, the third symbol for the
$(11000)^\infty$ process is highly informative and predictability
increases markedly.  

Corollary \ref{SyncPeriodic} applies here and, since $h_\mu = 0$, says
that the synchronization information $\SI$ is equal to $\TI$; and so,
we can directly interpret $\TI$ as the synchronization information. Table
\ref{Per5Sync} gives the values of the transient information $\TI$,
which are all different, indicating that an observer comes to
synchronize to the distinct templates differently.  Table
\ref{Per5Sync} also gives the average number of observations required
to synchronize.  From this table, we see that $\TI$ is not directly
proportional to the {\em number} of measurements to synchronize.
Rather, it is the total amount of {\em information} that must be
extracted to synchronize. 

In summary, this example shows that there are structural differences  
between different periodic processes of the same period.  The
transient information is able to capture these differences, while the
excess entropy is unable to.  Since many chaotic systems, for example,
are a combination of periodicity and randomness, one sees that the
transient information is useful in detecting synchronization to the
ordered component of such processes.

\subsection{Markovian Processes}
\label{Markovian_Example}

We now consider a simple Markovian process with a nonzero entropy rate
$h_\mu$. (The periodic systems of the previous section are Markovian,
but with $h_\mu = 0$.) In particular, we shall consider the golden
mean (GM) process, a Markov chain of order one.  

In terms of the sequences produced, the underlying golden mean
system produces all binary strings with no consecutive $0$s. The
probabilistic version --- the {\em golden mean process} --- generates
$0$s and $1$s with equal probability, except that once a $0$ is
generated, a $1$ is seen. One can write down a simple two-state Markov
chain for this process.  The GM process is so named because the
logarithm of the total number of allowed sequences grows with $L$ at
a rate given by the logarithm of the golden mean, $\phi =
\frac{1}{2}(1 + \sqrt{5})$. 

\begin{figure}[h]
\begin{center}
$\begin{array}{c}
\epsfxsize=3.0in
  \leavevmode
  \epsffile{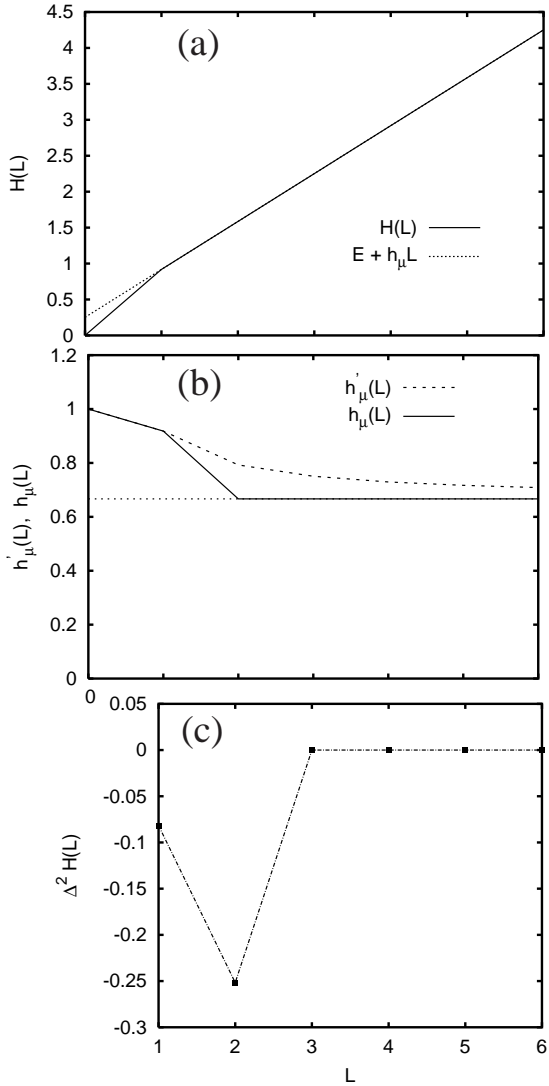} \\
\end{array}$
\end{center}
\vspace{-3mm}
\caption{(a) Entropy growth (solid line) for the golden mean process,
along with the asymptote $\EE + \hmu L$ (dashed line). (b) Entropy
convergence, both $\hmu(L)$ (solid line) and $\hmu^\prime (L)$ (dashed
line), for the same. (c) Predictability gain $\Delta^2 H(L)$ versus
sequence length. }
\label{GoldenMean}
\end{figure}

The various entropy convergence curves for the GM process are shown
in Fig.~\ref{GoldenMean}. The entropy rate of the GM process is
$\hmu = 2/3$ bits per symbol and the predictability is $\TP = 1/3$
bit per symbol. The convergence of $h_\mu(L)$ to $\hmu$ occurs at
sequence length $L=2$.  In other words, once the statistics over
all possible length-$2$ sequences are known, one gains no additional
predictability by keeping track of the occurrence of blocks of larger
length.  There is, however, a large predictability gain in going from
blocks of length $1$ to blocks of length $2$.  Observing that $00$ is
missing is the key observation that makes this system predictable.  
The predictability gain per symbol $\Delta^2 H(L)$ is shown in
Fig.~\ref{GoldenMean}(c). Note that the second measurement is more
informative than the first.

We find that $\EE \approx 0.2516$ bits, and $\TI = \EE$, which
can be easily deduced from the $H(L)$ versus $L$ graph
in Fig. \ref{GoldenMean}(a).  From
these small values for $\EE$ and $\TI$ one concludes that not much
historical information is needed to perform optimal prediction, nor is
there much uncertainty associated with synchronization. 

For this system we find that $H(1) \approx 0.9183$ bits. Plugging this
and our result for $h_\mu$ into Eq.~(\ref{EEMarkovian}), we see that
the expression for the excess entropy of a Markovian process is
verified.

The behavior shown in Fig.~\ref{GoldenMean} is typical for Markovian
processes.  For an order-$R$ Markovian process, the entropy density
estimates $h_\mu(L)$ will always converge exactly to $h_\mu$ by
$L=R$. This follows immediately from inserting Eq.
(\ref{OrderR.Markovian.def}) into the expression for $h_\mu(L)$,
Eq.~(\ref{DeltaHasCondEnt}).  Given this, we know that at $H(R) = \EE
+ h_\mu R$.  Solving this for $\EE$, we arrive at Eq.~(\ref{EEMarkovian}).

\subsection{Hidden Markov Processes I:  Complex Transient Structure}
\label{RRXOR_Example}

\begin{figure}[htb]
\begin{center}
$\begin{array}{c}
\epsfxsize=3.0in
\epsffile{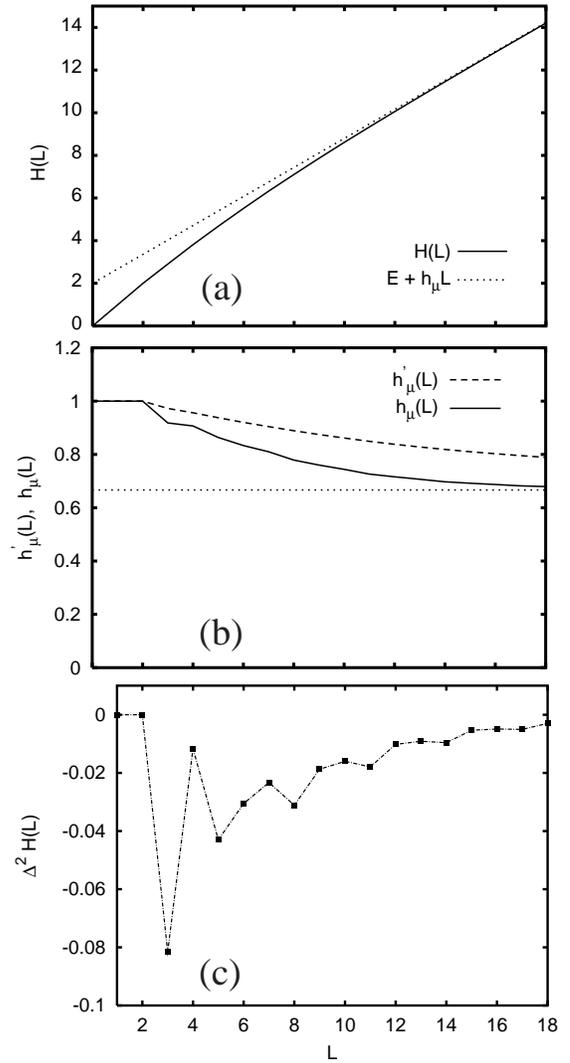} \\
\end{array}$
\vspace{-2mm}
\end{center}
\caption{(a) Entropy growth (solid line) for the random-random-XOR
process, along with the asymptote $\EE + \hmu L$ (dashed line). (b)
Entropy convergence, both $\hmu(L)$ (solid line) and $\hmu^\prime (L)$
(dashed line), for the same. (c) Predictability gain $\Delta^2 H(L)$
versus sequence length. }
\label{RRXOR}
\end{figure}

For our next three examples, we consider three different finitary
hidden Markov processes.  Each of these examples contains some
interesting surprises.  We begin by considering a process that
consists of two successive random symbols chosen to be $0$ or $1$ with
equal probability and a third symbol that is the logical Exclusive-OR
(XOR) of the two previous.  We call this the random-random-XOR (RRXOR)
process. The entropy growth and convergence plots are given in
Figs.~\ref{RRXOR}(a) and \ref{RRXOR}(b). 

\begin{figure}[h]
\begin{center}
\vspace{-6mm}
$\begin{array}{c}
\epsfxsize=3.0in
  \epsffile{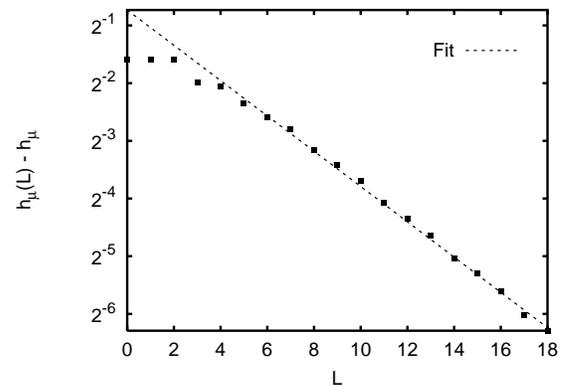} \\ [.2cm]
\end{array}$
\end{center}
\vspace{-2mm}
\caption{A least-squares fit (dashed line) to the exponential decay
of $\hmu(L)$ (squares) for the RRXOR process. }
\label{RRXORFit}
\end{figure}

The entropy rate $h_\mu$ is $2/3$ bits per symbol. To see this, note
that two out of every three symbols are completely random, while one
third of the symbols are determined by the previous two.  Note further
that the RRXOR process has the same $h_\mu$, and hence the same $\TP$,
as the GM process of the previous section. This serves as yet another
reminder that the entropy rate is not sufficient to distinguish the
structural properties of a source.  

At first blush, one might expect the entropy growth curve to reach its
asymptotic form at $L=3$, just as $H(L)$ did at $L=2$ for the golden
mean process.  However, Fig.~\ref{RRXOR}(a) shows that this is not the
case.  The reason that it does not converge exactly at $L=3$ is that
the RRXOR process is not Markovian; specifically, the observed sequences
of $0$'s and $1$'s are not finite-order Markovian. The RRXOR is a hidden
Markov process; its internal states are Markovian, but the observed
``states'' are not.  

Instead of converging exactly at finite $L$, the convergence of
$h_\mu(L)$ to $h_\mu$ is exponential:
\begin{equation}
  h_\mu(L) - h_\mu  = A 2^{-\gamma L} ~,
\end{equation}
where we find $A = .60 \pm 0.02$ and $\gamma = .306 \pm .004$.
This fit is illustrated in Fig.~\ref{RRXORFit}.

The excess entropy is $\EE = 2$ bits:  one needs to know which of the
four possible random symbol-pairs has occurred before one reaches a
condition of optimal predictability.  Thus, the process has $\log_2 4
= 2$ bits of memory.  However, the transient information is quite
large; $\TI \approx 9.43$ bit-symbols.  This indicates that the
process is difficult to synchronize to: Even after observing a large
number of symbols, there is still some uncertainty about which
internal, hidden state the process is in.  Nevertheless, the transient
information is finite. 

For this system, $H(1) = 1$. Using Eqs.~(\ref{EERelatedToGamma}) and
(\ref{TIRelatedToGamma}), we find $\EE_\gamma \approx 1.74$ bits and
$\TI_\gamma \approx 9.12$ bit-symbols. The differences from the near-exact
values above indicate the amount of deviation from a pure exponential
decay of $\hmu(L)$.

Intriguingly, the behavior of the predictability gain $\Delta^2 H(L)$
of Fig.~\ref{RRXOR}(c) shows strong hints of the structure of
the hidden Markov model that generates the observed sequences.  At
lengths $L = 1$ and $L = 2$ symbols are not informative at all:
$\Delta^2 H(L) = 0$.  This reflects the fact, given by the process's
definition, that two of the symbols are produced by fair coin flips. 
For larger $L$, note that $\Delta^2 H(L)$ shows oscillations of period
three.  The RRXOR hidden Markov process also has a period-$3$
structure: after the two random bits and the XOR bit, the hidden
Markov model always resets to the same state.  Recall, however, that
$\Delta^2 H(L)$ is formed from statistics over the observed symbols,
not the hidden states of the process.  Given this, it is somewhat
surprising that $\Delta^2 H(L)$ picks up the period-$3$ nature of the
transitions between hidden states.

\subsection{Hidden Markov Processes II: Measure Sofic Process}
\label{Even_Example}

We now consider another hidden Markov process: the {\em even process}
\cite{Crut91b}, a stochastic process whose support (the set of allowed
sequences) is a sofic system called the {\em even system}
\cite{Weis73}. The even system generates all binary strings consisting
of blocks of an even number of $1$s bounded by $0$s. Having observed a
process's sequences, we say that a word (finite sequence of symbols)
is {\em forbidden} if it never occurs.  A word is an {\em irreducible
forbidden word} if it contains no proper subwords which are themselves
forbidden words.  A system is {\em sofic} if its list of irreducible
forbidden words is infinite.  The even system is one such sofic
system, since its set $\{ 01^{2n+1}0, n = 0 , 1, \ldots \}$ of
irreducible forbidden words is infinite. Note that no finite-order
Markovian source can generate this or, for that matter, any other
strictly sofic system. The even process then associates probabilities
with each of the even system's sequences by choosing a $0$ or $1$ with
fair probability after generating either a $0$ or a pair of $1$s.  The
result is a {\em measure sofic process} --- a distribution over a
sofic systems sequences. Like the RRXOR process, the even system is not
Markovian, but a hidden Markov process.  

\begin{figure}[h]
\begin{center}
$\begin{array}{c}
\epsfxsize=3.0in
  \leavevmode
  \epsffile{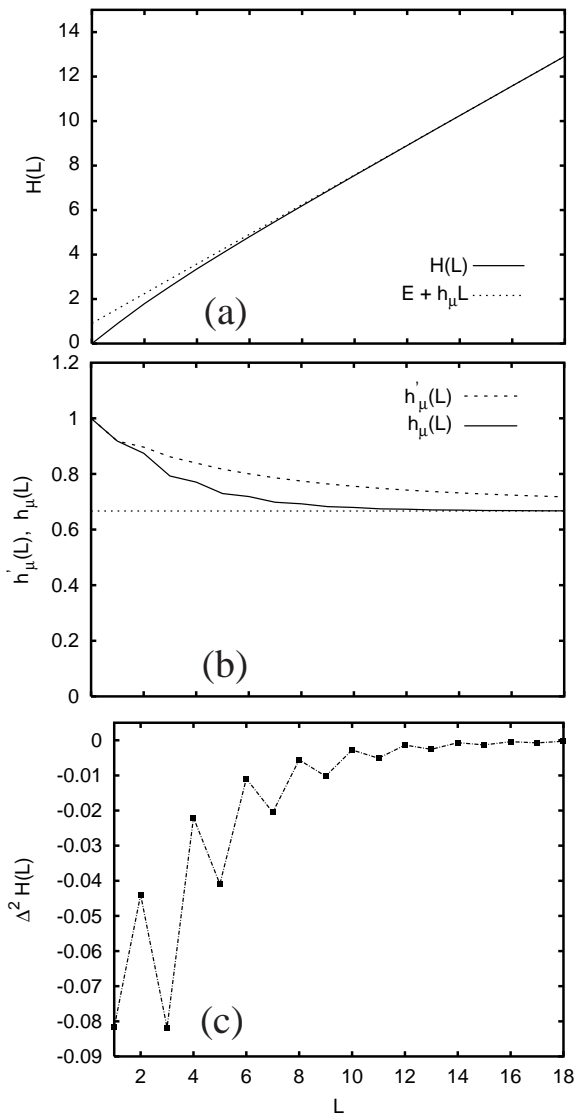} \\
\end{array}$
\end{center}
\vspace{-3mm}
\caption{(a) Entropy growth (solid line) for the even process, along
with the asymptote $\EE + \hmu L$ (dashed line). (b) Entropy
convergence, both $\hmu(L)$ (solid line) and $\hmu^\prime (L)$ (dashed
line), for the same. (c) Predictability gain $\Delta^2 H(L)$ versus
sequence length. 
  }
\label{Even}
\end{figure}

\begin{figure}[h]
\begin{center}
$\begin{array}{c}
\epsfxsize=3.0in
  \leavevmode
  \epsffile{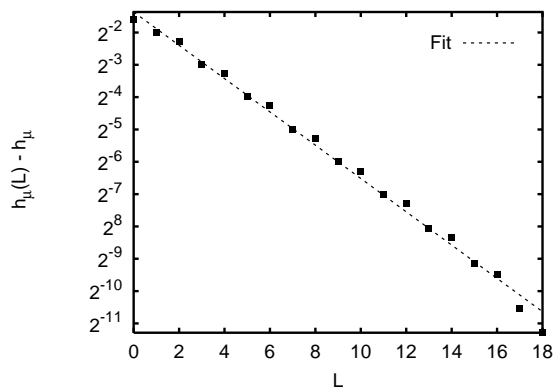} \\ [.2cm]
\end{array}$
\end{center}
\vspace{-.7cm}
\caption{A least-squares fit (dashed line) to the exponential decay
  of $\hmu(L)$ (squares) for the even process. }
\label{EvenFit}
\end{figure}

The various entropy convergence curves for the even process are shown 
in Fig.~\ref{Even}. The entropy rate of the even process is $\hmu =
2/3$ bits per symbol and the predictability $\TP$ is $1/3$ bits per
symbol. Note that these values are the same as those for the
RRXOR and GM processes, again emphasizing the poverty of $h_\mu$ as a
structural measure.  The convergence of $h_\mu(L)$ is exponential. A
fit to   
\begin{equation}
  h_\mu(L) - h_\mu  = A 2^{-\gamma L} ~,
\end{equation}
shown in Fig.~\ref{EvenFit}, yields $A = .388 \pm 0.019$ and
$\gamma = .501 \pm .007$.  We find that $\EE \approx 0.902$ bits. This
is the amount of storage required on average to hold the information
that a given observed $1$ is the ``even'' or ``odd'' symbol in a block
of $1$s. The transient information is $\TI \approx 3.03$ bit-symbols:
The even process is moderately difficult to synchronize to, although it
is much easier to synchronize to than the RRXOR process in the previous
example.  Since $H(1) \approx 0.918$, we find that $\EE_\gamma \approx
0.86$ and $\TI_\gamma \approx 2.92$, both of which agree well with the
values measured for $\EE$ and $\TI$. 

Again, the predictability gain per symbol $\Delta^2 H(L)$, shown in
Fig.~\ref{Even}(c), oscillates as it converges to zero.  The plot
indicates that odd-length measurement sequences are more informative
than even-length ones.  As in the RRXOR example, the oscillation of
$\Delta^2 H(L)$ provides a strong hint as to the underlying structure
of the hidden Markov process responsible for the observed sequences.
This process has two states and, thus, a strong period-$2$ component.
This periodic behavior in the hidden states is picked up in $\Delta^2
H(L)$, despite the fact that $\Delta^2 H(L)$ is based only on the
statistics of the observed, nonhidden symbols.

\vspace{-5mm}

\subsection{Hidden Markov Processes III: The Simple Nondeterministic Source}

We now consider a process known as the {\em simple nondeterministic 
source} (SNS).  This process was constructed to illustrate how
measurement distortion can contribute its own kind of apparent
structural complexity to a simple, but hidden, information source.
In particular, the SNS describes the process obtained via a
non-generating partition of the logistic map \cite{Crut92c}. 
For an introduction to issues of measurement-induced complexity
see Ref.~\cite{Crut92c}, and for a full mathematical treatment see
Ref.~\cite{Uppe97a}. Spatial versions of this class of hidden process
were introduced in Ref.~\cite{Crut91e} and analyzed from a computation
theoretic view in Ref.~\cite{Lind98a}.

\begin{figure}[h]
\begin{center}
\epsfxsize=3.0in
  \epsffile{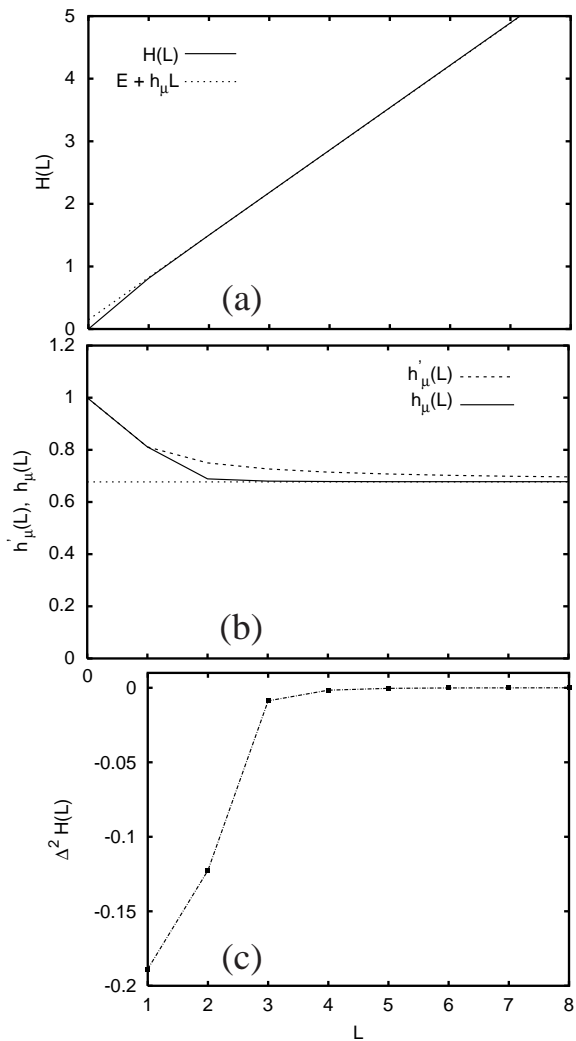}
\end{center}
\vspace{-6mm}
\caption{(a) Entropy growth (solid line) for the simple
nondeterministic source, along with the asymptote $\EE + \hmu L$
(dashed line). (b) Entropy convergence, both $\hmu(L)$ (solid line)
and $\hmu^\prime (L)$ (dashed line), for the same.  (c) Predictability 
gain $\Delta^2 H(L)$ versus sequence length. }
\label{Nondeterministic}
\end{figure}

The SNS, a hidden Markov process, generates symbol sequences as
follows. The system has three internal, hidden states:  $\bf A$, $\bf
B$, and $\bf C$. The observer, however, only sees the binary outputs
$0$ and $1$. The probabilities of generating the observed symbols,
when the process is in each of the internal states, are given by the
transition matrices $T^{(0)}$ and $T^{(1)}$, respectively: 
\begin{eqnarray}
T^{(1)} & = & \left [ \matrix{ 0 & 0 & 0 \cr
                      0 & 0 & 1/2 \cr
                      0 & 0 & 0 \cr } \right ]
\end{eqnarray}
and
\begin{eqnarray}
T^{(0)} & = & \left [ \matrix{ 1/2 & 1/2 & 0 \cr
                      0 & 1/2 & 0 \cr
                      1/2 & 1/2 & 0 \cr } \right ] ~.
\end{eqnarray}
The elements of the transition matrices are identified with the set of
internal states $\{ \bf A, \bf B, \bf C \}$ in the natural way.  For
example, $T_{23}^{(1)} = 1/2$ indicates that the probability of being
in state $\bf B$, producing a $1$, and making a transition to state $\bf C$
is $1/2$. 

Assuming that the observer knows the internal structure of the
process --- i.e., $T^{(0)}$ and $T^{(1)}$ --- then whenever a $1$ is
measured the observer knows that the internal state is $\bf C$.
However, for every $0$ measured after this, the observer becomes
and then remains uncertain as to whether the internal state is
$\bf A$ or $\bf B$. This also explains the label ``nondeterministic''
for this process: the measurement of $0$ does not {\em determine} the
internal state.  In contrast, all the previous examples we have
considered have been deterministic, in the sense that specifying the
output symbol determines the next internal state.

A central consequence of this nondeterminism is that the number of
{\em effective states} seen by an observer that attempts to reconstruct
the hidden process is infinite, even though the internal process is a
simple, three-state Markov chain \cite{Crut92c,Uppe97a}. The SNS is
arguably one of the simplest such examples for which this infinite-state
divergence occurs. 

\begin{figure}[h]
\begin{center}
  \epsfxsize=3.0in
  \epsffile{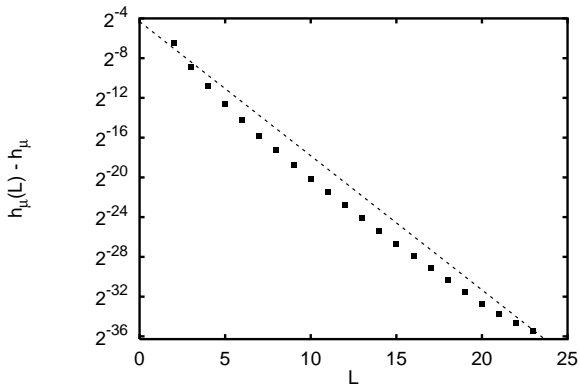}
\end{center}
\vspace{-4mm}
\caption{A semilog plot to test for an exponential decay of $\hmu(L)$ (squares).
  The latter are calculated exactly for sequences from $L=2$ to $L=25$.
  The dashed line represents an exponential decay.}
\label{NondeterministicExponentialFit}
\end{figure}

\begin{figure}[h]
\begin{center}
  \epsfxsize=3.0in
  \epsffile{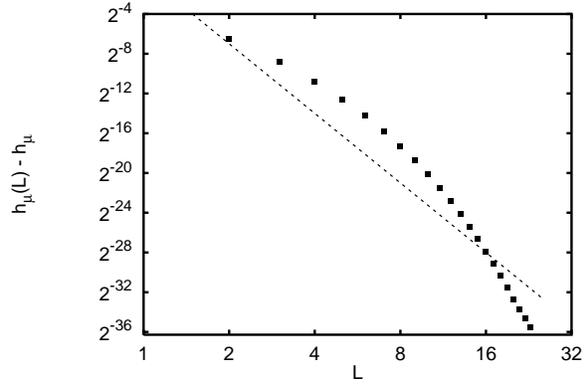}
\end{center}
\caption{A log-log plot to test for a power-law decay of $\hmu(L)$ (squares).
  The latter are calculated exactly for sequences from $L=2$ to $L=25$.
  The dashed line represents a power-law decay.}
\label{NondeterministicPowerLawFit}
\end{figure}


The various entropy convergence curves for the SNS process are shown in
Fig.~\ref{Nondeterministic}. The entropy rate, calculable analytically,
is $\hmu \approx 0.6778$ bits per symbol and the predictability is
$\TP \approx 0.3222$ bits per symbol. We find that $\EE \approx 0.147$
bits, there is not much mutual information between the past and future,
and $\TI \approx 0.175$ bit-symbols.  

Interestingly, the functional form of $h_\mu(L) - h_\mu$ is not clear.
An exponential decay
\begin{equation}
  h_\mu(L) - h_\mu \, = \, A 2^{-\gamma L} ~,
\end{equation}
is shown as the dashed line, with $A = 0.05$ and $\gamma = 1.35$, in
Fig.~\ref{NondeterministicExponentialFit}. One can also test a
power-law entropy decay of the form
\begin{equation}
  h_\mu(L) - h_\mu \, = \, cL^{-\alpha} \;.
\end{equation}
This is shown as the dashed line, with  $c = 1.0$ and $\alpha = 7.0$, in
Fig.~\ref{NondeterministicPowerLawFit}. Neither form is ideal: entropy
convergence is slower than exponential and faster a power law. Based on
Figs.~\ref{NondeterministicExponentialFit} and \ref{NondeterministicPowerLawFit}
one cannot infer a simple functional form for $h_\mu(L) - h_\mu$; perhaps it is
some version of a stretched exponential.

In short, the simple nondeterministic source has low predictability and
low apparent memory. Moreover, since $\TI$ is small, synchronizing to it
entails overcoming very little uncertainty. These would seem to be in
accord with the fact that one can write down a compact nondeterministic
representation for it that has only a few hidden states. However, to
perform optimal prediction, a deterministic representation is needed and for
the SNS that representation has an infinite number of states \cite{Crut92c}.
This degree of complexity is not suggested by the relatively small values for
the information theoretic measures of structure considered here. Thus,
relying only on information theoretic quantities, one is misled as to the
process's actual complexity. Nonetheless, the fact that entropy convergence
is not clearly exponential, in contrast to the even and RRXOR processes,
provides indirect evidence that the SNS is different from these other
finitary sources.  

\subsection{Aperiodic Infinitary Process}
\label{Morse-Thue_Example}

We now consider an infinite-memory process
that is aperiodic and has zero entropy rate. The {\em Thue-Morse (TM)
sequence} is the fixed point of the substitution $\sigma$ defined by:
\begin{eqnarray}
 \sigma(0) & = & 01 \;,  \\
 \sigma(1) & = & 10 \;.
\label{substitution.rule}
\end{eqnarray}
For example, starting from the initial string $s = 1$, the fifth
iterate in the TM sequence is:
\begin{equation}
  \sigma^5(s) = 10010110011010010110100110010110 \;.
\end{equation}
The {\em Thue-Morse language} $L_{TM}$ is the subset of all
words in the TM sequence:
\begin{equation}
  L_{TM} = {\rm sub} \left( \lim_{t \rightarrow \infty} \sigma^t (1)
\right)~, 
\end{equation}
where ${\rm sub} (s)$ gives all of the subwords in string $s$. The
{\em Thue-Morse process} is then given by assigning the natural
measure --- the frequency of occurrence in $\sigma^\infty (1)$ --- to
the words in $L_{TM}$. Unlike the previous three examples we have
considered, the Thue-Morse process is {\em not} generated by a
finitary hidden Markov process. In fact, there is no finite-state process
that can generate the Thue-Morse sequence.

\begin{figure}[h]
\begin{center}
\epsfxsize=3.0in
  \epsffile{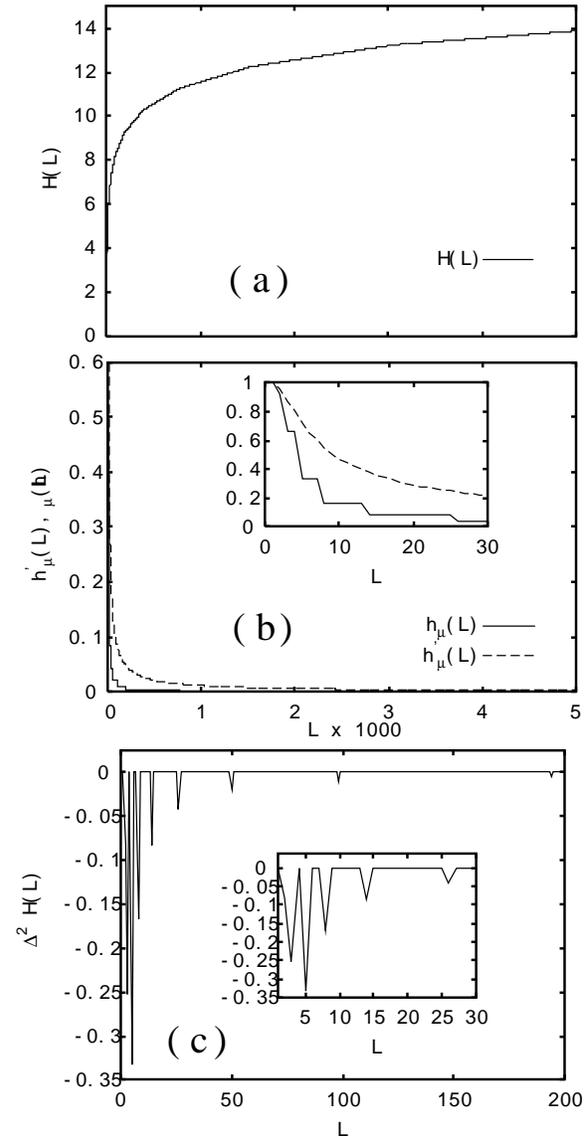}
\end{center}
\caption{(a) Entropy growth (solid line) for the Thue-Morse process. (b)
  Entropy convergence, both $\hmu(L)$ (solid line) and $\hmu^\prime (L)$
  (dashed line), for the same. In (a) and (b) sequence length goes up to
  $L = 5000$. (c) Predictability gain $\Delta^2 H(L)$ over small ranges
  of $L$.}
\label{MorseThue}
\end{figure}

\begin{figure}[h]
\begin{center}
\epsfxsize=3.0in
  \epsffile{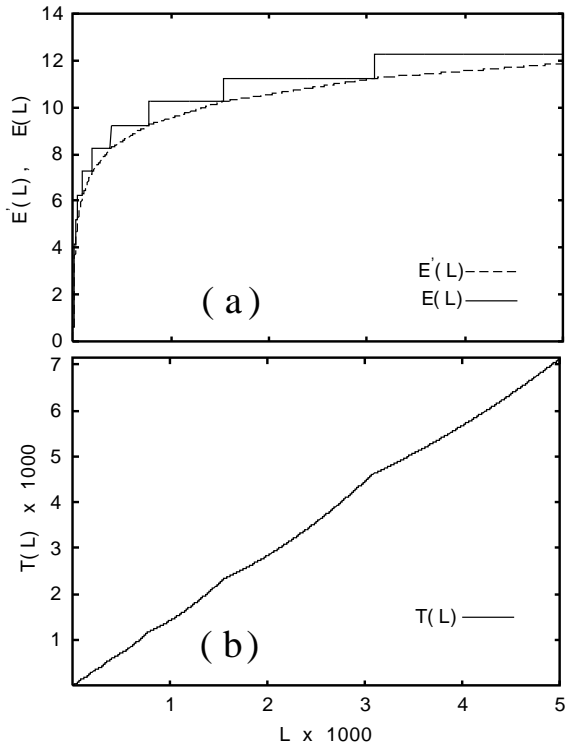}
\end{center}
\caption{(a) Excess entropy estimate divergence --- $\EE(L)$ (solid
line) and $\EE^\prime (L)$ (dashed line).   (b) Transient information
estimate divergence --- $\TI (L)$ (solid line). Note that the sequence
length goes up to $L = 5000$ and that both plots have large vertical
scales.} 
\label{MorseThueEandT}
\end{figure}

The various entropy convergence curves for the TM process are shown in
Figs.~\ref{MorseThue} and \ref{MorseThueEandT}.  These curves were
calculated using the results of Ref.~\cite{Bert94}, which show that:
\begin{eqnarray} 
  h_\mu(1) & = & 1 \;, \\
  h_\mu(2) & = & \log_2 3 - \frac{2}{3} \;, \\
  h_\mu(3) & = & \frac{2}{3} \;,
\end{eqnarray}
and, for $k \geq 1$:
\begin{equation}
  h_\mu(L) = \left\{
    \begin{array}{ll}  
      4 / (3 \cdot 2^k), & {\rm if} \;\;  2^k + 1 \leq L-1 \leq 3 \cdot
      2^{k-1}\;  \\
      2 / (3 \cdot 2^k), & {\rm if} \;\; 3 \cdot 2^{k-1} + 1 \leq L-1
      \leq 2^{k+1}  \\
      \end{array} \right. \;.
\end{equation}

From this, one concludes that $h_\mu = 0$ and that the entropy-rate
estimates converge according to a power law: $\hmu(L) \propto 1/L$.
Thus, the total entropy grows logarithmically: $H(L) \propto \l2 L$;
as shown in Fig.~\ref{MorseThue}(a). Despite the slow convergence  
to $\hmu = 0$, the predictability is high: $\TP = 1$ bit per symbol.
Each measurement gives the maximal amount of information about the
nonrandom part of the process.

Nonetheless, the excess entropy diverges; $\EE(L) \propto \l2 L$,
indicating an infinite-memory process. (See
Fig.~\ref{MorseThueEandT}(a).)  This can be also be inferred from
Fig.~\ref{MorseThue}(a), where $\EE$ is simply the height of the
$H(L)$ curve, since $h_\mu = 0$.  Finally, the transient information
estimate $\TI(L)$ also diverges, linearly, as shown in
Fig.~\ref{MorseThueEandT}(b). This linear divergence is explained by
looking at Eq.~(\ref{TI.as.integral.by.parts}).  If one substitutes
$h_\mu(L) \sim L^{-1}$ and $h_\mu = 0$ into the expression there for
$\TI$, the linear divergence follows immediately.  

It is clear from Fig.~\ref{MorseThue}(c) that there are long sequences
of measurements that are uninformative. These are punctuated
occasionally with isolated symbols that do improve
predictability. These occur at sequence lengths $L_i = 3 \times
2^{i-3} + 2, ~i = 3, 4, 5, \ldots$.  To determine why 
$\Delta^2 H(L)$ behaves in this manner requires a computation
theoretic approach, such as that given in Ref.~\cite{Crut90} for the
symbolic dynamics produced at the period-doubling accumulation point
of the logistic map. For another similar approach, see
Ref.~\cite{Wang99a}. 

We conclude this section by noting that, based on our results and
those of several other authors \cite{Gras86,Freu96a,Bert94,Gram94},
this $1/L$ entropy convergence is typical of aperiodic sequences
generated by substitutions rules like those of
Eq.~(\ref{substitution.rule}).  Moreover, Freund, Ebeling, and
Rateitschat have given an argument for why this entropy convergence
form is characteristic of aperiodic sequences \cite{Freu96a}. 

\subsection{Other Infinitary Processes}

Before concluding this section, we review the results of several other 
investigations of entropy convergence. Sz\'epfalusy and Gy\"orgyi
\cite{Szep86} found that $h_\mu(L) - h_\mu \sim L^{-\alpha}$ with
$\alpha \geq 5/2$ for a class of one-dimensional intermittent maps.   
Thus, this class consists of finitary processes. However,
for a different model of an intermittent process, Freund found a
similar decay form, but with $\alpha \approx 0.492$ \cite{Freu96a}.
Examining temporal-block sequences in elementary one-dimensional
cellular automata, Grassberger \cite{Gras86} also found a power law
decay, with $\alpha = 0.6 \pm 0.1$ for rules 30 and 45 and $\alpha =
1.0 \pm 0.1$ for rule 120. These are examples of infinitary processes.

A number of researchers have examined entropy convergence for written
texts---such as The Bible, Grimms' Tales, Moby Dick, the gnuplot manual,
and Gleick's popular book ``Chaos''
\cite{Ebel94c,Ebel97b,Ebel91,Anis94,Ebel95d}. The picture that
emerges is that entropy convergence can be fit to a power law
$h_\mu(L) - h_\mu \sim L^{\alpha}$ with $\alpha$ ranging from $0.4$ to
$0.6$. Interestingly, for a Beethoven sonata an exponent of
$\alpha \approx .75$ has been found \cite{Anis94}. Again, these
results indicate infinitary processes.

Recently, Nemenman \cite{Neme00a} and Bialek, Nemenman, and Tishby
\cite{Bial00a} have found power-law convergences for different
one-dimensional Ising models. For long-range coupling, where the
coupling constants decay as the inverse lattice separation, they found
an $\alpha$ of $0.5$. They also examined an Ising model with
short-range interactions, but in which the coupling constant changes
every $400,000$ sites within a lattice of $10^9$ spins. The coupling
constant was drawn from a Gaussian distribution with zero mean. For this
system they found a power-law decay with an exponent of $\alpha = 1$.   

\subsection{Summary of Examples}

For comparison, Table \ref{SummaryOfExamples} collects the
various analytical and numerical estimates of the information
theoretic quantities for the preceding examples we analyzed.

\begin{table}[t]
\begin{center}
\begin{tabular}{|l|c c c c c|}
Process & $\hmu$ & $\TP$ & $\gamma$ & $\EE$ & $\TI$ \\
\hline
Fair Coin         & $1$     & $0$     &         & $0$     & $0$ \\
Biased Coin       & $0.881$ & $0.119$ &         & $0$     & $0$ \\
Period-$16$       & $0$     & $1$     &         & $4$     & $16.6135$ \\
$(11000)^\infty$  & $0$     & $1$     &         & $\l2 5$ & $4.073$ \\
$(10000)^\infty$  & $0$     & $1$     &         & $\l2 5$ & $5.273$ \\
$(10101)^\infty$  & $0$     & $1$     &         & $\l2 5$ & $4.873$ \\
Golden Mean       & $2/3$   & $1/3$   &         & $0.252$ & $0.252$ \\
Even              & $2/3$   & $1/3$   & $0.501$ & $0.902$ & $3.03$ \\
Random-Random-XOR & $2/3$   & $1/3$   & $0.306$ & $2$     & $9.43$ \\
Nondeterministic  & $0.678$ & $0.322$ & $1.35^{\ast}$ & $0.147$ & $0.175$ \\
Thue-Morse        & $0$     & $1$     &  & $\propto \log L$ & $\propto L$ \\
\end{tabular}
\caption{Summary of Examples. $^\ast$This can also be fit to a 
  power law, $L^{-\alpha}$ with $\alpha \approx 7$.
  }
\end{center}
\label{SummaryOfExamples}
\end{table}

\section{Applications and Implications}
\label{Applications}

Being cognizant of various types of entropy convergence, of different 
classes of process, and of how to quantitatively distinguish between
them is useful general knowledge. To this end, we reviewed
information-theoretic quantities, introduced a new one, the transient 
information, and put forth a unified framework for relating them all
in terms of discrete derivatives and integrals.  Then, in the
preceding section, we analyzed a number of examples.  We return now to
the set of questions posed in the introduction: How can we
untangle different sources of apparent randomness? In particular,
what happens to our estimates of the entropy rate if we ignore a
process's structure? 

Addressing these questions is the task of this last section. Here 
we show that there are direct and empirically important consequences 
for ignoring structural properties. We consider several different
questions:
\begin{enumerate}
\item What happens when an observer ignores entropy-rate convergence?
\item What happens when the process's apparent memory is ignored?
\item On the one hand, what happens if the observer ignores synchronization? 
\item On the other hand, what happens if the observer assumes it is
synchronized to the process? 
\end{enumerate}

The answers, given below, show that ignoring a process's structural
properties leads to a range of misleading inferences about randomness
and organization. In addition to highlighting the negative
consequences, we also comment on the fact that the associated problems
can be alleviated to some extent, even in cases where data is
limited.

\subsection{Disorder as the Price of Ignorance}

The first two questions are closely related and rather straightforward 
to answer.  The preceding sections defined several different
quantities --- $\hmu$, $\TP$, $\EE$, and $\TI$ --- that measure
randomness, predictability, memory, synchronization, and other
features of a process. For the most part, these are asymptotic
quantities in the sense that they involve the behavior of the function
$H(L)$ in the $L \rightarrow \infty$ limit. Thus, their exact
empirical estimation demands that an infinite number of measurements
(for accurate estimates of sequence probabilities) of infinitely long
sequences be made.  Obviously, other than by analytic means, it is not
possible to exactly calculate such quantities. Exact, $L \rightarrow
\infty$ results are known for only a few special systems which are
analytically tractable.

This leads one to ask, even when sequence probabilities are accurately
known, how well can these various source properties be estimated at
finite $L$?  What errors are introduced and are these errors related
in any way? 

The simplest such question, the first one listed above, arises when
one attempts to estimate source randomness $h_\mu$ via the
approximation $h_\mu(L)$.  Stopping the estimate at finite $L$ gives
one a rate $\hmu(L)$ that is larger than the actual rate $\hmu$.  That
is, the source appears {\em more random} if we ignore correlations
between variables separated by more than $L$ steps.  This observation
follows directly from the definitions of $h_\mu$ and $h_\mu(L)$.
However, it turns out that this form of overestimation of $h_\mu$ is
related to the excess entropy $\EE$.  We shall see that there is a
quantitative trade-off between randomness and memory.

Assume an observer makes measurements of a process with entropy rate 
$\hmu$ and excess entropy $\EE > 0$.  Recall the definition,
Eq.~(\ref{ent.def}), of the entropy rate. Using this definition to
estimate $\hmu$ is tantamount to assuming that $\EE = 0$ --- see the 
dashed line $\hmu L$ in Fig.~\ref{HvsL}. But, by assumption, $\EE
> 0$. Thus, at a given $L$, we can ask what the entropy estimate
$\hmu^\prime (L) = H(L)/L$ is.  Lemma
\ref{EntropyRateConvergeFromAbove} established that $\hmu^\prime (L) >
\hmu$.  But by how much more?  This is answered in a straightforward
way by the following proposition.

\begin{proposition}
When the observer is synchronized to the process,
\begin{equation}
h_\mu^\prime (L) - \hmu = \frac{\EE}{L} ~.
\end{equation}
\label{StructureToRandom}
\end{proposition}

\begin{figure}[tbp]
\epsfxsize=3.0in
\begin{center}
\leavevmode
\epsffile{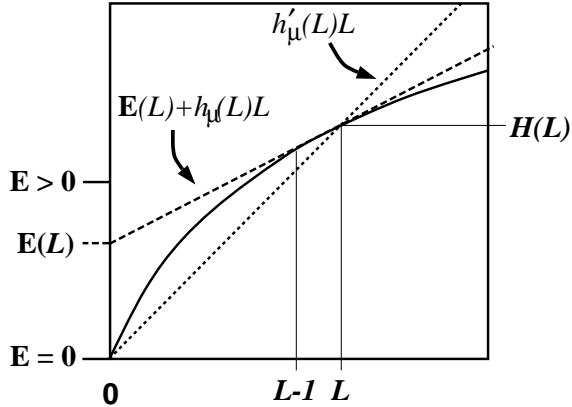}
\end{center}
\vspace{2mm}
\caption{Ignored memory is converted to randomness: Illustration of how
ignoring memory, in this case implicitly assuming $\EE = 0$ as
Eq.~(\ref{ent.def}) implies, when actually $\EE > 0$, leads to an
overestimate $\hmu^\prime (L)$ of the actual entropy rate $\hmu$.} 
\label{MemoryToDisoder}
\end{figure}

\noindent {\em Proof:}
The claim follows immediately from the graphical construction given in
Fig.~\ref{MemoryToDisoder}. Saying that the observer is synchronized
to the source means using an $L$ such that $H(L) = \EE + \hmu
L$. Thus, 
\begin{eqnarray}
\nonumber
\hmu^\prime (L) & = & \frac{H(L)}{L} \\
  & = & \frac{\EE + \hmu L}{L} ~.
\end{eqnarray}
Eq.~(\ref{StructureToRandom}) follows directly. $\Box$

In this way, $\EE$ bits of memory are converted into additional,
apparent randomness.  The process appears more random due to the
observer ignoring one of its structural properties.

One can object to this estimate: Typically one does not know the
process's properties (e.g., $\EE$ and $\hmu$) and so even these must
be estimated.  Thus, expressing the estimator $h_\mu^\prime$ in terms
of the asymptotic quantities $\EE$ and $\hmu$ may not be that
useful.  However, $\EE^\prime (L)$ in Eq.~(\ref{EEPrimeEstimate}) is a
non-asymptotic, $L$-dependent estimator of memory.  Namely,
$\EE^\prime (L)$ is a measure of the mutual information between two
halves of an $L$-block.  Using this estimator we can restate
Proposition \ref{StructureToRandom}. 

\begin{proposition}$\;$
\begin{equation}
\hmu^\prime (L) - \hmu(L) \geq \frac{\EE^\prime (L)}{L} ~.
\end{equation}
\end{proposition}

\noindent {\em Proof:}
Again, see Fig.~\ref{MemoryToDisoder}.
Appealing to the monotonicity and convexity of $H(L)$, the
monotonicity of $\hmu(L)$, and Lemma
\ref{EntropyRateConvergeFromAbove}, we can rewrite the definition
\begin{equation}
\EE(L) = H(L) - \hmu(L) L \;,
\end{equation}
as
\begin{equation}
\frac{H(L)}{L} - \hmu(L) = \frac{\EE(L)}{L} ~.
\end{equation}
Since $\EE^\prime (L)$ is bounded above by $\EE(L)$ by Lemma
\ref{EEConvergenceBounds}, we have
\begin{equation}
\hmu^\prime (L) - \hmu(L) \geq \frac{\EE^\prime (L)}{L} ~.
\end{equation}
which directly proves the claim. $\Box$

This result establishes how $\hmu(L)$ lower bounds $\hmu^\prime (L)$,
as indicated by Lemma \ref{EntropyRateConvergeFromAbove}.  In
particular, it emphasizes that their difference is controlled by the
excess entropy, a measure of memory. 

Although $\EE$ is an $L$-asymptotic quantity, the error $\EE/L$ in the
entropy-rate estimate dominates at small $L$. Moreover, being
restricted to small $L$ is typical of experimental situations with
limited data or in which drift is present. That is, one cannot
reliably estimate the $L$-block probabilities $\Pr (s^L)$ at large $L$
due to the exponential growth in their number or the nonstationarity
of block probabilities, respectively.

\subsection{Predictability and Instantaneous Synchronization}

Conversely, if one assumes a fixed amount of memory $\EE$, we shall
see that this leads to an underestimate of the entropy rate $h_\mu$. 
Assuming a fixed excess entropy is not something that one would be
likely to do in the particular setting here, in which an observer
empirically measures entropy density and related quantities from
observed symbol sequences.  In a more general modeling setting,
however, one always runs the risk of over-fitting and, in so doing,
``projecting'' some particular structure---such as, additional
memory capacity---onto the system.  Assuming a
fixed, nonzero value for the excess entropy is, in an abstract sense,
an example of over-fitting.  Given this, we ask, What is the
consequence of assuming a fixed value for $\EE$? 

\begin{figure}[tbp]
\epsfxsize=3.0in
\begin{center}
\leavevmode
\epsffile{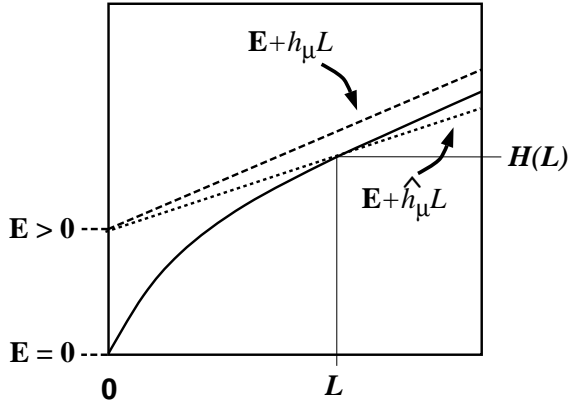}
\end{center}
\vspace{2mm}
\caption{Assumed synchronization converted to false predictability:
  Schematic illustration of how assuming one is synchronized to a
process, leads to an underestimate $\widehat{\hmu}$ for a source with 
excess entropy $\EE > 0$ and entropy rate $\hmu$.} 
\label{InstantSyncOrder}
\end{figure}

Equivalently, what happens if the observer assumes that it is
synchronized to the process at some finite $L$, implying that
$H(L) = \EE + h_\mu L$?  The geometric construction for this scenario
is given in Fig.~\ref{InstantSyncOrder}.  In effect the source is
erroneously considered to be a completely observable Markovian
process in which, as we have seen, $H(L)$ converges to its asymptotic
form exactly at some finite $L$.  If the observer then uses
Eq.~(\ref{EEMarkovian}) to estimate $h_\mu$ using its assumed value
for $\EE$, one arrives at the estimator $\widehat{\hmu}$ where
\begin{equation}
\widehat{\hmu} \, \equiv \, \frac{H(L) - \EE}{L} \, \neq \hmu \;.
\label{widehat.hmu.def}
\end{equation} 
At a given $L$ the effect is
that the observer considers the source to have a larger $\EE$ than it
actually has at that $L$. The line $\EE + \widehat{\hmu} L$ is fixed
at $\EE$ when that intercept should be lower. The result, easily gleaned
from Fig.~\ref{InstantSyncOrder}, is that the entropy rate $\hmu$ is
underestimated as $\widehat{\hmu}$.  In other words, the source
appears more predictable than it actually is. 

\begin{proposition}$\;$
An observer monitors a process with excess entropy $\EE > 0$. If the
observer assumes it is synchronized when it is not, then
\begin{equation}
\widehat{\hmu} \leq \hmu \;.
\end{equation}
\end{proposition}

\noindent {\em Proof:}
From Fig.~\ref{InstantSyncOrder} or Eq.~(\ref{widehat.hmu.def}), one
sees that 
\begin{equation}
\widehat{\hmu} = \frac{H(L) - \EE}{L} ~.
\end{equation}
The observer is assuming that it is seeing $H(L) = \EE +
\widehat{\hmu} L$.  But since $H(L) \leq \EE + \hmu L$, we have that
\begin{equation}
\EE + \widehat{\hmu} L \leq \EE + \hmu L ~,
\end{equation}
and so $\widehat{\hmu} \leq \hmu$.  $\Box$

\subsection{Assumed Synchronization Implies Reduced Apparent Memory}

\begin{figure}[tbp]
\epsfxsize=3.0in
\begin{center}
\leavevmode
\epsffile{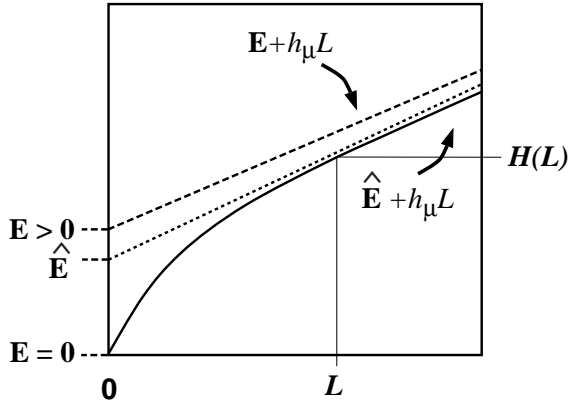}
\end{center}
\vspace{2mm}
\caption{Assumed synchronization leads to less apparent memory:
Schematic illustration of how assuming synchronization to a source, in
this case implicitly assuming $H(L) = \EE + \hmu L$, leads to an
underestimate $\widehat{\EE}$ of the actual memory $\EE > 0$.}
\label{SyncLessMemory}
\end{figure}

In addition to analyzing the effects on the apparent entropy rate due
to assuming synchronization, we can ask a complementary question:
What are the effects on estimates of the apparent memory
$\widehat{\EE}$?  Figure \ref{SyncLessMemory} illustrates this
situation. 

If, at a given $L$, we approximate the entropy rate estimate
$H(L) - H(L-1)$ by the true entropy $\hmu$, then the offset between
the asymptote and $H(L)$ is simply
\begin{equation}
\Delta \EE = \EE + \hmu L - H(L) ~.
\end{equation}
Translating this back to the original we have a reduced apparent
memory $\widehat{\EE} \leq \EE$ of
\begin{equation}
\widehat{\EE} = H(L) - \hmu L ~.
\end{equation}
In fact, since the estimated entropy rate is larger than $\hmu$,
the reduction in apparent memory is even larger.

Thus, assuming synchronization, in the sense that $h_\mu(L) = h_\mu$, leads
one to underestimate the apparent memory, as measured by the excess entropy
$\EE$.  

\section{Conclusion}
\label{Conclusions}

Looking back, we have introduced a variety of information theoretic
measures of a process's randomness and a variety of structural
properties.  Along the way, we put forth a new quantity, the transient
information $\TI$.  One of the central results of this work is
contained in Theorem \ref{SyncTheorem}, where we proved that $\TI$
is related to the total state-uncertainty experienced while
synchronizing to a Markov process.  

We also calculated these information theoretic quantities for a range
of differently structured processes.  A natural question, then, is:
To what extent does this information theoretic approach allow us to
distinguish between processes that are structured in fundamentally
different ways?

\subsection{Process Classification}

To summarize our results from Section \ref{Examples_Section}, we now
give a rough classification of several types of information source
based on the quantities studied here. Similar, although coarser,
classifications have been put forth by Sz\'epfalusy \cite{Szep89a},
Ebeling \cite{Ebel97b}, and Crutchfield \cite{Crut92c}.  

First, we have the zero entropy rate, asymptotically predictable processes.
\begin{enumerate}
\item {\em Periodic processes}: For period-$p$ processes, $H(L)$ becomes a
constant and $\hmu(L)$ vanishes for $L \geq p$.
\item {\em Aperiodic processes}: These are infinitary processes, since they
need, in a crude sense, an infinite amount of memory to maintain their
aperiodicity. Having $\hmu = 0$, they cannot be aperiodic by virtue of an
internal source of randomness. $\TI$ diverges, indicating that one is never
fully synchronized.
\end{enumerate}

Then we have the positive entropy rate, irreducibly unpredictable processes.
\begin{enumerate}
\item {\em Memoryless processes}: For these, $H(L)$ scales as $\hmu L$
and $\hmu(L)$ converges immediately to $\hmu$. We have $\EE = 0$ and
$\TI = 0$. Independent, identically distributed (IID) processes are
examples of this class. They have no temporal memory and no
structural complexity. 
\item {\em Finitary processes}: In this class $H(L)$ scales as
$\EE + \hmu L$. The entropy density $\hmu(L)$ typically converges
exponentially to $\hmu$. We have $0 < \EE < \infty$ and $\TI >
0$. Ref.~\cite{Junc79} established a useful connection between
information and ergodic theories for this class: finite $\EE$ means
that a process is weak Bernoulli.  Within the finitary class further
structural distinctions are possible: 
\begin{enumerate}
\item {\em Markov processes}: The basic property of
Markovian sources is that one synchronizes to them exactly at some
finite block length $L$.  For these processes, the effective states
can be taken to be single symbols or symbol blocks of some finite
length.  Once that length of sequence has been parsed,
the observer is synchronized and can then optimally predict the
process.  
\item {\em Deterministic hidden Markov processes}: These processes are
characterized by an exponential convergence of $h_\mu(L)$, in contrast
to the exact convergence at finite $L$ exhibited by a Markov process.
Depending on the transition structure of the hidden states, these
processes can have relatively large values for the excess entropy and
transient information.  Within this broad class of hidden Markov
processes lies the interesting case of a measure sofic process --- a
system whose support set contains an infinite list of irreducible
forbidden words.  In a limited sense, these systems have an infinite
memory that keeps track of the (infinite) list of
irreducible forbidden words.  Nevertheless, the measure sofic process
considered here, the even process, had finite $\EE$ and $\TI$.  As
noted above, the behavior of $\Delta^2 H(L)$ for these processes seems
to provide strong hints of the structure of the hidden state
transitions responsible for the infinite memory. In particular, we
find that $\Delta^2 H(L)$ oscillates with a periodicity given by the
periodic structure of the transitions between hidden states.

\item {\em Nondeterministic hidden Markov processes}:  
It would appear that this class of process may not be overtly
different from other finitary hidden Markov processes.  However, the
example we considered, the simple nondeterministic source, showed a
markedly different entropy convergence behavior than the other hidden
Markov examples.  

\end{enumerate}
\item {\em Infinitary Sources}:
At this point in time, this remains a catch-all category of processes
--- those falling outside the finitary classes. These include, for
example, various context-free languages, such as positive-entropy-rate
variations on the Thue-Morse
process and other stochastic analogues from higher up the Chomsky
hierarchy.  Presumably, within the infinitary sources there are many
interesting structural distinctions waiting to be discovered; some
analogous to the automata-architectural distinctions recognized by
discrete computation theory \cite{Hopc79} and some distinctions
related to the nature of the measure over the infinite sequences.
\end{enumerate}

The ultimate goal of this type of classification would be an
amalgamation of the structural distinctions made in the Chomsky
hierarchy of computation theory \cite{Hopc79} and statistical
categories found in the ergodic theory hierarchy of stochastic
processes \cite{Lebo73a}. 

Recent work by Nemenman \cite{Neme00a} and Bialek, Nemenman, and Tishby
\cite{Bial00a} may be a helpful step in this direction. In
Refs.~\cite{Bial00a,Neme00a} they show that the excess entropy --- the
``predictive information'' in their parlance --- is, in some circumstances,
related to the number of parameters in the model producing the
process.  However, this result holds in a slightly different context
than ours.  Rather than using histograms of larger and larger
variables blocks, they consider a procedure in which an observer is
trying to learn a distribution through successive samplings.

\subsection{Inferring Models from Finite Resources}

In Section \ref{Applications} we considered various trade-offs between
finite-$L$ estimates of the excess entropy $\EE$, the transient
information $\TI$, and the entropy rate $h_\mu$.  In particular, we
have shown that not taking one or another into account leads one to
systematically {\em over-} or {\em underestimate} a source's entropy
rate $\hmu$.  For example, there can be an inadvertent conversion
of ignored memory into apparent randomness.  The magnitude of this
effect is proportional to the difference between source memory
and the upper bound on memory that the observer can estimate. In a
complementary way, one can inadvertently convert assumed memory into
false predictability. One eventually comes to see that a process's
structural features must be accounted for, even if one's focus is only
on an apparently simpler question of (say) how random a process is
\cite{Ford83}.

\subsection{Future Directions}

We conclude by mentioning some important open questions and suggesting
several directions for future research.  First, at a number of points
we have referred to ``structure'', without actually defining it. Is
there a better, more systematic, and principled approach for
determining the structure of an information source than the pure
information-theoretic one just outlined?  Refs.~\cite{Shal98a} and
\cite{Crut92c}, for example, argue that {\em computational mechanics}
is a viable approach to quantifying source structure and the patterns
produced by information sources. They show that the $\epsilon$-machine
representation used there captures all of a source's structure. Thus,
one natural question is how one can determine entropy convergence
behavior given a process's $\epsilon$-machine. 

Second, it would be helpful to make a direct connection between the
source characterization developed here --- in terms of average source
properties measured by $\hmu$, $\EE$, $\TI$, and $\TP$ ---
and the difficulty of estimating these quantities and of inferring
models of the sources. Analyzing the computational complexity of these
two problems is the domain of computational learning
theory \cite{Kear94a,Vapn95a}.

Third, establishing that the source entropy rate $\hmu$ is a metric
invariant is one of the hallmarks of ergodic and dynamical systems
theories \cite{Kolm58,Kolm59,Sina59}. What status do $\EE$ and $\TI$
hold in the same setting?

Finally, there is, of course, the question of how the information
theoretic approach to structure outlined here can be extended to more
than one dimension. There has been some preliminary work in this direction
\cite{Gras86,Andr00,Arno96,Lind91a,Lind97,Sicl97a,Pias00a}; however,
many questions remain.  One of the central difficulties is that,
unlike in one dimension where the various expressions for the excess
entropy are equivalent, they yield different results when extended to
two dimensions \cite{Feld98c}.  Careful definitions and distinct
interpretations of the different forms of two-dimensional excess entropy
and related quantities will have to be given in order to develop a
useful, fully two-dimensional approach to pattern and structure. Our
hope is that the preceding development is sufficiently clear and thorough
that it can serve as a firm foundation for an information theory of
structure in higher-dimensional processes. 

\section*{Acknowledgments}

The authors thank Kristian Lindgren for helpful discussions and
Robert Haslinger and Cosma Shalizi for comments on the manuscript.
The cost-of-amnesia view of $\EE$ is due to Dan Upper. This work was
supported at the Santa Fe Institute under the Computation, Dynamics,
and Inference Program via SFI's core grants from the National Science
and MacArthur Foundations. Direct support was provided from DARPA
contract F30602-00-2-0583 and AFOSR via NSF grant PHY-9970158. DPF
thanks the Department of Physics and Astronomy at the University of
Maine, Orono, for their hospitality during the summer of 2000, when
part of this work was completed.  


\bibliography{spin}

\appendix

\section{Proofs}
\label{PropProofs}

\subsection{Prop.~\ref{DeltaHisInfoGain}}
\label{ProofDeltaHisInfoGain}

\noindent  {\bf Proposition \ref{DeltaHisInfoGain}}:
$\Delta H(L) = {\cal D}[ {\rm Pr}(s^L) || {\rm Pr}(s^{L-1}) ]$. \\

\noindent {\em Proof}: By direct calculation we have the following.
\begin{eqnarray}
{\cal D} [ {\rm Pr} (s^L) || {\rm Pr} (s^{L-1}) ] 
  & = &  \sum_{ \{ s^L \} } {\rm Pr} (s^L)
  \l2 \frac{ {\rm Pr} (s^L) } { {\rm Pr} (s^{L-1}) } \label{gain}\\    
  & & \hspace{-3cm} = \sum_{\{ s^L \}} {\rm Pr} (s^L) \l2 {\rm Pr}
  (s^L) - \nonumber \\ 
  &  & \hspace{-1cm} \sum_{\{ s^{L-1} \}} \sum_{\{ s_{L-1} \}}
  {\rm Pr} (s^L) \l2 {\rm Pr} (s^{L-1}) \\
  & & \hspace{-3cm}=  H(L) - \sum_{\{ s^{L-1} \}} \l2 {\rm Pr} (s^{L-1})
  \sum_{\{ s_{L-1} \}} {\rm Pr} (s^L) \\
  & &\hspace{-3cm} =  H(L) - H(L\!-\!1) \;,
\label{GainDerivative}
\end{eqnarray}
since ${\rm Pr} (s^{L-1}) = \sum_{\{ s_{L-1} \}} {\rm Pr}(s^L)$.  $\Box$

\subsection{Prop.~\ref{DeltaSqrHisCondlInfoGain}}
\label{ProofDeltaSqrHisCondlInfoGain}

\noindent {\bf Proposition \ref{DeltaSqrHisCondlInfoGain}}: 

$\Delta^2 H(L) = - {\cal D} [ {\rm Pr}(s_{L-1}|s^{L-2}) || {\rm
Pr}(s_{L-2}|s^{L-3}) ]$.\\ 

\noindent {\em Proof:}
By the expressions for the second discrete derivative,
Eq.~(\ref{Curvature}) and Eq.~(\ref{gain}), we have:
\begin{eqnarray}
  \Delta^2 H(L) & = & \Delta H(L) - \Delta H(L\!-\!1) \\ \nonumber 
  & = & -\sum_{\{ s^L \}} {\rm Pr}(s^L) \log_2 {\rm Pr}(s_{L-1}|s^{L-1})\\ 
  & & + \sum_{\{ s^{L-1} \}} {\rm Pr}(s^{L-1}) \log_2 {\rm Pr}(s_{L-2} |
  s^{L-2}) \\  
  & = & -\sum_{\{ s^L \}} {\rm Pr}(s^L) \log_2
  \frac{{\rm Pr}(s_{L-1}|s^{L-1}) }{{\rm Pr}(s_{L-2}|s^{L-2})} 
\label{lindgren.form} \\
  & = & - {\cal D} [ {\rm Pr}(s_{L-1}|s^{L-1})
  || {\rm Pr}(s_{L-2}|s^{L-2}) ] \;. 
\end{eqnarray}
$\Box$

\subsection{Prop.~\ref{TotalPredisRedundancy}}
\label{ProofTotalPredisRedundancy}

\noindent {\bf Proposition \ref{TotalPredisRedundancy}}:
$-{\TP} = {\bf R}$. \\

\noindent {\em Proof:}
We write the sum of Eq.~(\ref{TotalPredictability}) and use the
anti-differentiation formula Eq.~(\ref{Integration}) to get:
\begin{equation}
\sum_{L=1}^M \Delta^2 H(L) = \Delta H(M)- \Delta H(0) \;.
\end{equation}
Since $\lim_{L \rightarrow \infty} \Delta H(L) = h_\mu$ and since we
have defined $\Delta H(0) = \l2 |\Abet| $, it follows immediately that 
\begin{eqnarray}
  - {\TP} & \, = \,& \lim_{M \rightarrow \infty} [ \Delta H(0) - \Delta
  H(M) ] \\ 
  & \, = \, & \l2 | \Abet | - \hmu \;,
\end{eqnarray}
which is ${\bf R}$ by Eq.~(\ref{Redundancy}).$\Box$

\subsection{Prop.~\ref{TotalPredisCorrInfo}}
\label{ProofTotalPredisCorrInfo}

\noindent {\bf Proposition \ref{TotalPredisCorrInfo}}: $\TP = -
\sum_{L=2}^\infty (L\!-\!1) \Delta^3 H(L)$. \\ 

\noindent {\em Proof:}
We write Eq.~(\ref{total.predictability.as.average.corr.info}) as a
partial sum as follows:
\begin{eqnarray}
  \sum_{L=2}^\infty (L\!-\!1) \Delta^3 H(L) \, & = \, \nonumber \\ 
    & & \hspace{-3cm}  \lim_{M \rightarrow \infty} \left[
    \sum_{L=2}^M L \Delta^3 H(L) - \sum_{L=2}^M \Delta^3 H(L) \right]
    \;.   
\end{eqnarray}
We use Eqs.~(\ref{IntegrationByParts}) and (\ref{Integration}) on the
first and second terms on the right-hand side and obtain, after
simplifying: 
\begin{eqnarray}
   - \sum_{L=2}^\infty (L\!-\!1) \Delta^3 H(L) &\, = & \nonumber \\
   & & \hspace{-3cm} - \lim_{M \rightarrow \infty}
   \left[ M \Delta^2 H(M) - \sum_{L=1}^M \Delta^2 H(L) \right] \;.
\end{eqnarray}
From the definition of $\TP$, Eq.~(\ref{TotalPredictability}), and
since we assume that $\TP$ is finite,
$\lim_{M \rightarrow \infty} L \Delta^2 H(L) = 0$.
From this we see immediately that 
\begin{equation}
  - \sum_{L=2}^\infty (L\!-\!1) \Delta^3 H(L)
  = \sum_{L=1}^\infty \Delta^2 H(L) \equiv \TP \;. 
\end{equation} $\Box$

\subsection{Prop.~\ref{TotalExcessEntropy}}
\label{ProofTotalExcessEntropy}

\noindent {\bf Proposition \ref{TotalExcessEntropy}}: 
$\EE = -\sum_{L=2}^\infty (L\!-\!1) \Delta^2 H(L) $. \\ 

\noindent {\em Proof:} Writing the right-hand side of the above
equation as a partial sum, and then using the integration-by-parts
formula Eq.~(\ref{IntegrationByParts}) we obtain, after some algebra: 
\begin{eqnarray} 
-\sum_{L=2}^\infty (L\!-\!1) \Delta^2 H(L) \, & = &  \, \nonumber \\
 & & \hspace{-3cm} \lim_{L \rightarrow \infty}  \left\{ -M \Delta H(M)
   \,+\, \sum_{L=1}^M  \Delta H(L) \right \}\;.
\end{eqnarray}
Recalling that $\Delta H(L) = h_\mu(L)$ and that $h_\mu(M) \rightarrow
\hmu$ in the $M \rightarrow \infty$ limit, we see at once that
\begin{equation}
-\sum_{L=2}^\infty (L\!-\!1) \Delta^2 H(L) \, = \, 
 \sum_{L=1}^\infty [\hmu(L) - \hmu] \, \equiv \, \EE \;. 
\end{equation}
The last equality follows from the definition of $\EE$,
Eq.~(\ref{E.def}). 
$\Box$

\subsection{Prop.~\ref{EEfromEntropyGrowth}}
\label{ProofEEfromEntropyGrowth}

\noindent {\bf Proposition \ref{EEfromEntropyGrowth}}:
$\EE = \lim_{L \rightarrow \infty} [ H(L) - \hmu L ]$. \\

\noindent {\em Proof}: Writing out the partial sum of the infinite
sum in Eq.~(\ref{E.def}) and evaluating it using the integration
formula, Eq. (\ref{Integration}):
\begin{equation}
  \sum_{L=1}^M [ \Delta H(L) - \hmu ] = H(M) - H(0) - \hmu M \;.  
\end{equation}
Since $H(0) \equiv 0$, it then follows immediately that
\begin{equation}
  \EE = \lim_{M \rightarrow \infty} [ H(M) - \hmu M ] \;.
\end{equation} 
Since, by Eq.~\ref{IntrinsicRedundancy}, the left-hand side is
${\bf R}(L)$, the proof is complete.
$\Box$

\subsection{Prop.~\ref{EandMI_Proposition}}
\label{ProofEandMI}

\noindent {\bf Proposition \ref{EandMI}}: $\EE = I [ \future ; \past
]$. \\

\noindent {\em Proof}:
We rewrite the definition so that we can use the
finite-$L$ forms of various entropies:
\begin{equation}
I [ \future ; \past ]
  \equiv \lim_{L \rightarrow \infty} I [\future^L ;\past^L] \;.
\end{equation}
We begin with the definition of mutual information,
Eq.~(\ref{mutual.info.ent.diff}), which expresses $I$ as the difference
between two entropies:
\begin{equation}
I [ \future^L ; \past^L ]
  = H[ \future^L ] - H[ \future^L | \past^L ] \;.
\label{MIasConditionalMarginal}
\end{equation}
Recall that $H[\future^L] = H(L)$.

Using the conditional entropy chain rule \cite{Cove91} we have
\begin{eqnarray}
H[ \future^L | \past^L ]
  & = & H[ S_0, S_1, \ldots, S_{L-1} | S_{-L}, \ldots, S_{-1} ] \\
  & = & \sum_{i=0}^{L-1} H[ S_i | S_{-L} S_{-L+1} \cdots S_{i-1} ] \;.
\end{eqnarray}

Putting these together we have
\begin{eqnarray}
I [ \future ; \past ] &\, = \,& \nonumber \\
    & & \hspace{-2cm} \lim_{L \rightarrow \infty}
  \left[ H(L) -  \sum_{i=0}^{L-1} H[ S_i | S_{-L} S_{-L+1} \cdots
  S_{i-1} ] \right] \;.
\end{eqnarray}
In the $L \rightarrow \infty$ limit, each term in the summand is equal
to $h_\mu$.  Thus, we see that
\begin{equation}
I [ \future ; \past] \, = \, \lim_{L \rightarrow \infty}
  [ H(L) - L \hmu ] \;,
\end{equation}
which is $\EE$ by Prop. \ref{EEfromEntropyGrowth}.
$\Box$

\subsection{Lemma \ref{EntropyRateConvergeFromAbove}}
\label{ProofEntropyRateConvergeFromAbove}

{\bf Lemma \ref{EntropyRateConvergeFromAbove}}:
$\hmu^\prime (L) \geq \hmu(L) \geq \hmu$. \\

\noindent {\em Proof:}
We prove the right inequality first. Since conditioning reduces entropy,
\begin{equation}
\hmu(L) \geq \hmu(L^\prime) \;,\; \forall L > L^\prime \;.
\label{decreasing}
\end{equation}
Now, recall that 
\begin{equation}
\lim_{L \rightarrow \infty} \hmu(L) = \hmu \;.
\end{equation}
Since, by Eq.~(\ref{decreasing}), the $h_\mu(L)$'s are nonincreasing
as $L$ increases, it follows that $\hmu(L) \geq \hmu$.

We now prove the left inequality in the proposition.  
\begin{eqnarray}
\nonumber
\hmu^\prime(L) & \equiv & \frac{H(L)}{L} \\
   & = & \frac{1}{L} \sum_{i=1}^L H[S_i|S_{i-1} S_{i-2} \cdots S_i] \;.
\end{eqnarray}
For all $i<L$, 
\begin{equation}
H[S_i|S_{i-1} S_{i-2} \cdots S_i]
  \geq H[S_L|S_{L-1} S_{L-2} \cdots S_1] \;.
\end{equation}
Thus, 
\begin{eqnarray}
\hmu^\prime(L) & \geq & \frac{1}{L} 
  \sum_{i=1}^L H[S_L|S_{L-1} S_{L-2} \cdots S_1] \\
  & = & \frac{1}{L} L   H[S_L|S_{L-1} S_{L-2} \cdots S_1] \\
  & = & h_\mu(L)  \;.
\end{eqnarray}
$\Box$

\subsection{Lemma \ref{EEConvergenceBounds}}
\label{ProofEEConvergenceBounds}

{\bf Lemma \ref{EEConvergenceBounds}}: $\EE^\prime(L) \leq \EE (L)
\leq \EE$. \\ 

\noindent {\em Proof:}
We first prove the right inequality. Recall that
\begin{equation}
\EE^\prime(L) \equiv H(L) - L \hmu(L)
  = \sum_{M=1}^L( \hmu(M) - \hmu(L) ) \;.
\end{equation}
Since $M \geq L$ for all terms in the summand, all elements of
the sum are positive. Now, the excess entropy is defined as
\begin{equation}
\EE \equiv \lim_{L\rightarrow \infty}
  \sum_{M=1}^L( \hmu(M) - \hmu(L) ) \;.
\end{equation}
Thus, $\EE (L)$ is the partial sum of the above term. Since all
terms in the sum are non-negative, it follows immediately that
the partial sum $\EE (L)$ is less than the infinite sum $\EE$.  

We now prove the left inequality. Using stationarity,
\begin{equation}
\EE^\prime(L) = 2H(L/2) - H(L) \;.
\end{equation}
Recall that for odd $L$, we defined $\EE^\prime(L) = \EE^\prime(L-1)$.
To prove the left inequality, it will suffice to show that:
\begin{equation}
2H(L/2) - H(L) \leq H(L) - L \hmu(L) \;.
\end{equation}
Rearranging, we have:
\begin{equation}
2H(L/2) \leq 2H(L) - L \hmu(L) \;.
\end{equation}
By the concavity of $H(L)$, $2H(L/2) \geq H(L)$, and thus the above
equation becomes:
\begin{equation}
H(L) \leq 2H(L) - L \hmu(L) \;.
\end{equation}
Rearranging again, we see that we need to show:
\begin{equation}
H(L) \geq L \hmu(L) \;. 
\label{obvious}
\end{equation}
That this equation is true can be seen geometrically by inspecting
Fig.~\ref{MemoryToDisoder}. Note that the inequality is saturated
if and only if the process is independent identically distributed.

To verify Eq.~(\ref{obvious}) algebraically, we use the chain rule
on the left hand side and obtain:
\begin{equation}
\sum_{M=1}^L H[S_M|S_{M-1}S_{M-2} \cdots S_1] \geq L \hmu(L) \;.
\end{equation}
But,
\begin{eqnarray}
\nonumber
\sum_{M=1}^L H[S_M|S_{M-1}S_{M-2} \cdots S_1]
  & & \nonumber \\
 & & \hspace{-2cm} \geq \, \sum_{M=1}^L H[S_L|S_{L-1}S_{L-2} \cdots S_1] \\
  &  & \hspace{-2cm} =  L \hmu(L) \;. 
\end{eqnarray}
Thus, Eq.~(\ref{obvious}) is true, and the proof is complete. 
$\Box$

\section{Exponential Convergence to the Entropy Rate}
\label{ExponentialDecayAppendix} 

It was claimed in the main text that $h_\mu(L) - h_\mu$ often vanishes
exponentially fast for finitary sources. Why is this behavior so common?
There are several ways to argue for the ubiquity of exponential entropy
convergence.

First, note that if $\Delta^2 H(L)$ converges to $0$ exponentially
fast, then $h_\mu(L) = \Delta H(L)$ must also converge exponentially
fast.  Then, a direct calculation shows that $\Delta^2 H(L) \leq
I(L)$, where $I(L)$ is the mutual information between two variables
separated by $L$ symbols.  Now, the two-variable mutual information is
related to the two-variable correlation function $C(L)$.  In
particular, $I(L) \propto C^2(L)$.  This result was first shown for
binary sequences by Li \cite{Li90} and later generalized to larger
alphabets by Herzel and Grosse \cite{Herz95}. As a result, if the
correlations decay exponentially, then the two-symbol mutual
information decays exponentially. This, in turn, allows one to
conclude that the entropy-rate estimate converges exponentially
and so $\EE$ is finite.

The conclusion from these observations is that exponential convergence
of correlation functions implies the exponential convergence of the
entropy rate.  However, this only transfers the convergence question from
entropy rates to correlation functions. So {\em why} is it that
correlation functions typically decay exponentially? There are
several answers to this question.

Mathematically, many stochastic processes can be re-expressed as
one-dimensional spin models; see, e.g., Ref.~\cite{Beck93}. Thus, we
expect that what is typical for spin systems will also be typical for
the more general stochastic processes of interest to us here. In a
one-dimensional statistical mechanical model with finite
interaction strengths, one can always express the partition function
as an infinite product of transfer matrices.  The correlation function
between two spins $L$ lattice sites apart is proportional to
$(\lambda_0/\lambda_1)^L$, where $\lambda_0$ is the largest eigenvalue
of the transfer matrix and $\lambda_1$ the second largest eigenvalue.
The Perron-Frobenius theorem guarantees that the largest eigenvalue is
nondegenerate, thus establishing the exponential decay of the correlation
function. This result is standard; see, for example, Ref.~\cite{Binn92}.  

Physically, in a spin system the sum of the correlation functions
yields the magnetic susceptibility $\chi$. The exponential decay of the
correlation function thus ensures that $\chi$ is finite.  Hence, away
from a critical point, where we expect finite response functions such as
$\chi$, we also expect exponentially decaying correlation functions
--- or at least correlations that decay faster than $1/L$.  

Mathematically, it has been shown that, under a fairly wide range of 
circumstances, a statistical mechanical system with an analytic
partition function necessarily has correlation functions that decay
exponentially \cite{Lebo73b}. Unlike the Perron-Frobenius transfer
matrix argument, the results in Ref.~\cite{Lebo73b} hold for systems
in more than one dimension.

\section{Proof of the Synchronization Information Theorem }
\label{SyncMarkovProof}

We begin by restating the theorem:

{\bf Theorem} \ref{SyncTheorem}: If the source is order-$R$ Markovian, then 
\begin{equation}
\SI = \TI + \frac{1}{2} R(R\!+\!1) \hmu \;.
\label{SyncTheoremEquation.2}
\end{equation}

\noindent {\em Proof:} Since the transition probabilities are
normalized, $T$ is a stochastic matrix; $\sum_b T_{ab} = 1$. The
eigenvector corresponding to the eigenvalue $1$ shall be denoted
by $\pi$ and is normalized in probability; 
\begin{equation}
  \sum_a \pi_a T_{ab} \, = \, \pi_b, \;\;\; \sum_a \pi_a = 1 \;.
\end{equation} 
As is well-known, $\pi_a$ gives the asymptotic probability of the state
$A \in {\cal V}$.  Equivalently, in terms of the $R$-blocks,
\begin{equation}
  \pi_A \, = \, {\rm Pr}( \varphi^{-1}(A) ) \;.
\label{pi.symbols}
\end{equation}
Or, simply
\begin{equation}
  \pi_A \, = \, {\rm Pr}(s^R) \;, 
\end{equation}
where $s^R$ is understood to correspond to the $A$th state. 

Initially, before any measurements are made, we assume our
distribution over $\cal V$ is given by $\pi$;
\begin{equation}
   {\rm Pr}( {\cal V} | \lambda, {\rm M} ) = \pi \;,
\end{equation}
where $\lambda$ is the empty string.   Hence, ${\cal H}(0) =
H\{\pi\}$.  Equivalently, it follows from
Eqs.~(\ref{state.Rblock.equiv}) and (\ref{pi.symbols}) that
\begin{equation}
  {\cal H}(0) = H(R) \;.  
\end{equation}
If we observe a particular symbol $s^\prime_1$, we now know
that the process must be in one of the states that correspond to
symbol blocks whose first symbol is $s^\prime_1$.  We denote this set
of states by:
\begin{equation}
  {\cal V}_{s^\prime_1} \equiv \{ \varphi({s^\prime_1} s_2 \cdots s_R):
  s_i \in \Abet, 2 < i < R \} \;.   
\end{equation}
Likewise, after we've observed the particular length $L$ sequence
${s^\prime}^L$, $L<R$, we know that the process must be in one of the
states that corresponds to an $R$-symbol block whose first $L$ symbols
are ${s^\prime}^L$;
\begin{eqnarray}
  \hspace{-1cm}{\cal V}_{{s^\prime}^L} \, &\equiv &\, \{ \varphi({s^\prime}^L s_{L+1}
  s_{L+2} \cdots s_R): \nonumber \\
   & & \hspace{.5cm}  s_i \in \Abet, L+1 < i < R \} \;, ~L \leq R \;.
\label{subset.V.def}
\end{eqnarray}

The following properties of ${\cal V}_{{s^\prime}^L}$ follow immediately from
the definition, Eq.~(\ref{subset.V.def}):
\begin{equation}
   {\cal V}_{{s}^L} \subset {\cal V} \;,
\end{equation}
\begin{equation}
    {\cal V}_{{s}^L} \bigcap {\cal V}_{{s^\prime}^L} \,= \, \emptyset
    \;\; {\rm if \;and \;only \;if} \; s^L \neq {s^\prime}^L \;,
\label{empty.intersection}
\end{equation}
and,
\begin{equation}
  \bigcup_{s^L} {\cal V}_{s^L} \, = \, {\cal V}, \;.  
\end{equation}
Thus, the set of $L$-blocks $\{ s^L \}$ induces a partition of the set
of states $\{ {\cal V} \}$.  For a given $L$ there are at most
$\Abet^L$ sets ${\cal V}_{s^L}$, each of which is a proper subset of
${\cal V}$.  (There are exactly $\Abet^L$ subsets of ${\cal V}$ if and
only if there are no forbidden sequences.) The set ${\cal V}_{s^L}$ has
at most $\Abet^{R-L}$ elements.  So, as more and more symbols are
observed --- i.e., as $L$ grows --- the subsets ${\cal V}_{s^L}$ of
${\cal V}$ become more and more refined.  For the Markovian case
considered here, eventually enough symbols will be observed so that we
know with probability $1$ the state of the process.  Since the
Markovian states are in a one-to-one relation with the $R$-blocks, we
are guaranteed to know the state with certainty after $R$ symbols have
been observed.  Hence, ${\cal H}(R) = 0$.  Observing subsequent
symbols will not add to the state uncertainty since each observation
uniquely determines the subsequent state.  Thus, ${\cal H}(L) = 0$ for
$L \geq R$. 

For $L<R$, the distribution over the Markovian states $v \in {\cal V}$
is given by:
\begin{equation}
  {\rm Pr}(v|s^L, {\cal M} ) \, = \, \frac{ \pi^{s^L}}{{\rm
  Pr}(s^L)}\;, 
\label{prob.V.L}
\end{equation}
where $\pi^{s^L}$ is a vector whose $|{\cal V}|$ components are
given by:
\begin{equation}
  (\pi^{s^L})_v \, = \, \left\{ 
    \begin{array}{ll}  
      \pi_v, & {\rm if} \; v \in {\cal V}_{s^L} \\
      0, & {\rm otherwise}  \\
      \end{array} \right. \;.
\label{pi.sL.def}
\end{equation}
We are interested in calculating ${\cal H}(L)$, the average
state-uncertainty after observing $L$ symbols. In order to perform
this calculation, the following two properties of $\pi^{s^L}$ will be
necessary.  

First, for fixed $s^L$, observe that summing $(\pi^{s^L})_v$ over its
components $v$ results in ${\rm Pr}(s^L)$, the probability of that
particular $s^L$.  This follows from the definition of $(\pi^{s^L})$,
Eq.~(\ref{pi.sL.def}): 
\begin{eqnarray}
  \sum_{\{ v \}} (\pi^{s^L})_{v \in {\cal V}} & \, = \, & \sum_{ v \in {\cal
      V}_{s^L} } \pi_v \\ 
  & \, = \, & \sum_{ \{ s^R : \varphi(s^R) \in {\cal V}_{ s^L} \}} {\rm
    Pr}(s^R)\\ 
  & \, = \, & \sum_{ \{ s_{L+1} s_{L+2} \cdots s_R \}} {\rm Pr}(s^R) \\
  & \, = \, & {\rm Pr}(s^L) \;.
\label{property.1}
\end{eqnarray}
Hence, ${\rm Pr}({\cal V}|s^L,{\cal M})$ as given in
Eq.~(\ref{prob.V.L}) is normalized over $s^L$.  


Second, notice that $(\pi^{s^L})_v$ has only one nonzero entry for
fixed $L$ and fixed  state $A$.  This follows from noting that the
particular state $A \in {\cal V}$ is associated with a particular
$R$-block $\varphi^{-1}(A)$.  More formally, suppose that
$(\pi^{s^L})_v$ has a nonzero entry for two different $L$-blocks, say
$s^L$ and ${s^\prime}^L$; 
\begin{equation}
  (\pi^{s^L})_v \, = \, (\pi^{ {s^\prime}^L})_v \, > \, 0, \;\; s^L
  \neq {s^\prime}^L \;.
\end{equation}
Then, by Eq.~(\ref{pi.sL.def}), it follows that:
\begin{equation}
  A \in {\cal V}_{s^L},\;\; {\rm and} \; v \in {\cal V}_{ {s^\prime}^L} \;,
\end{equation}
which, in turn, implies that:
\begin{equation}
{\cal V}_{s^L} \cap {\cal V}_{ {s^\prime}^L} \neq \emptyset
~{\rm and}~ s^L \neq {s^\prime}^L \;.
\end{equation}
This last equation contradicts Eq.~(\ref{empty.intersection}).  Thus,
the original proposition must be true:  $(\pi^{s^L})_v$ has only one
nonzero entry --- namely $\pi_v$ --- for all possible $s^L$'s.


We are now ready to complete our calculation of ${\cal H}(L)$.
Plugging Eq.~(\ref{prob.V.L}) into Eq.~(\ref{script.H.def}) and
simplifying slightly, we have:
\begin{eqnarray}
  {\cal H}(L) \, &=& - \, \sum_{\{ s^L \}} \sum_{v \in {\cal V}}
  (\pi^{s^L})_v  \log_2
  (\pi^{s^L})_v  + \nonumber \\ 
  & &  \hspace{1cm} \sum_{\{ s^L \}} \sum_{v \in {\cal V}} (\pi^{s^L})_v
  \log_2 {\rm Pr}(s^L)  \;.    
\end{eqnarray}
Parenthetically, we note that ${\cal H}(L)$ is the information gain:
${\cal H}(L) = {\cal D} [ \pi^{s^L} || \Pr (s^L) ]$.
By Eq.~(\ref{property.1}), we can perform the sum over $v$ in the
second term on the right-hand side of the above equation, and we
obtain the entropy of an $L$-block, $H(L)$.

To evaluate the first term on the right-hand side, recall that
$(\pi^{s^L})_v$ has only one nonzero entry for fixed $L$ and fixed
$v$.  Using this, we see that
\begin{eqnarray}
\nonumber
  \sum_{s^L} & & \left( - \sum_{v \in {\cal V}} (\pi^{s^L})_v \log_2 (\pi^{s^L})_v \right) \\
  & = & - \sum_{v \in {\cal V}} \pi_v \log_2 \pi_v \\
  & = & H[ \pi ] \\
  & = & H(R) \;.
\end{eqnarray}
Thus, it follows that:
\begin{equation}
  {\cal H}(L) = \left\{
    \begin{array}{ll}
      H(R) - H(L) & {\rm if}\; 0 \leq L \leq R \\
      0 & {\rm if}\; L > R \\ 
    \end{array} \right. \;.
\end{equation}

We now have an expression for ${\cal H}(L)$ in terms of $H(L)$, and we
finish the proof with a direct calculation.  Looking at
Eq.~(\ref{SyncTheoremEquation.2}), one sees that it will suffice to
show that:
\begin{equation}
  \sum_{L=0}^\infty{\cal H}(L) - \TI \, = \, \frac{h_\mu}{2}R(R\!+\!1)
  \;. 
\label{ProofStep}
\end{equation}
By assumption, the process is order-$R$ Markovian.  This
implies that ${\cal H}(L) = 0$ and $H(L) = \EE + h_\mu L$ for all $L
\geq R$.  As a result of this latter equation, the summand of
the infinite sum that defines $\TI$, Eq.~(\ref{T.def}), is zero for all
$L \geq R$. That is, the last nonzero contribution to the sum comes at
$L=R-1$.  As a result, the left-hand side of Eq.~(\ref{ProofStep}) can
be written as:
\begin{eqnarray}
  \sum_{L=0}^\infty{\cal H}(L) - \TI \, & = & \,  \nonumber \\
 & & \hspace{-2cm} \sum_{L=0}^{R-1} \left[ H(R) - H(L) -\EE - h_\mu L + H(L) \right] \\
  & & \hspace{-1cm} = \, \sum_{L=0}^{R-1} \left[ H(R) -\EE - h_\mu L
  \right] \;.  
\end{eqnarray}

But $H(R) = \EE + h_\mu R$, since we assume that synchronization occurs
at $L=R$.  Plugging this into the above equation, we have:
\begin{eqnarray}
 \sum_{L=0}^\infty{\cal H}(L) - \TI &\, = \,& \sum_{L=0}^{R-1} \left[
 \EE + h_\mu R - \EE - h_\mu L  \right] \\
 & \, = & \, h_\mu \sum_{L=0}^{R-1} ( R - L ) \\
  & \, = & \, h_\mu (R^2 - \frac{1}{2} R(R-1) \\
   & \, = & \, \frac{h_\mu}{2}R(R\!+\!1) \;. 
\end{eqnarray}
This last equation is Eq.~(\ref{ProofStep}), thus completing the
proof. $\Box$. 

\end{multicols}

\end{document}